\documentclass[jcp,twocolumn,superscriptaddress,floatfix]{revtex4}
\usepackage{dcolumn} 
\usepackage{bm} 
\usepackage{mathtools}
\usepackage{braket}
\usepackage[version=3]{mhchem}
\usepackage{xspace}
\usepackage{textcomp}
\usepackage{gensymb}
\usepackage{cleveref}

\begin{document}

%%%%%%%%%%%%%%%%%%%%%%%%%%%%%%%%%%%%%%%%%%%%%%%%%%%%%%%%%%%%%%%%%%%%%%
% Title
%%%%%%%%%%%%%%%%%%%%%%%%%%%%%%%%%%%%%%%%%%%%%%%%%%%%%%%%%%%%%%%%%%%%%%
\title{Analytic energy gradients for variational two-electron reduced-density matrix methods within the density-fitting approximation}

%%%%%%%%%%%%%%%%%%%%%%%%%%%%%%%%%%%%%%%%%%%%%%%%%%%%%%%%%%%%%%%%%%%%%%
% Authors
%%%%%%%%%%%%%%%%%%%%%%%%%%%%%%%%%%%%%%%%%%%%%%%%%%%%%%%%%%%%%%%%%%%%%%
\author{J.~Wayne~Mullinax}
\affiliation{
             Department of Chemistry and Biochemistry,
             Florida State University,
             Tallahassee, FL 32306}
\author{Evgeny~Epifanovsky}
\affiliation{
             Q-Chem, Inc.,
             6601 Owens Drive, Suite 105,
             Pleasanton, CA 94588}
\author{Gergely~Gidofalvi}
\affiliation{
             Department of Chemistry and Biochemistry,
             Gonzaga University, 
             Spokane, Washington  99258}
\author{A.~Eugene~DePrince~III}
\affiliation{
             Department of Chemistry and Biochemistry,
             Florida State University,
             Tallahassee, FL 32306}
\email{deprince@chem.fsu.edu}

%%%%%%%%%%%%%%%%%%%%%%%%%%%%%%%%%%%%%%%%%%%%%%%%%%%%%%%%%%%%%%%%%%%%%%
% Commands
%%%%%%%%%%%%%%%%%%%%%%%%%%%%%%%%%%%%%%%%%%%%%%%%%%%%%%%%%%%%%%%%%%%%%%
\newcommand{\kcal}{kcal mol$^{-1}$\xspace}
\newcommand{\req}{\ensuremath{r_{\rm e}}\xspace}
\newcommand{\aeq}{\ensuremath{\theta_{\rm e}}\xspace}
\newcommand{\weq}{\ensuremath{\omega_{\rm e}}\xspace}
\newcommand{\cm}{cm$^{-1}$\xspace}
\newcolumntype{.}{D{.}{.}{-1}}

%\begin{document}

%\begin{tocentry}
%    \includegraphics{v2rdm_df_gradients_toc}
%\end{tocentry}

%%%%%%%%%%%%%%%%%%%%%%%%%%%%%%%%%%%%%%%%%%%%%%%%%%%%%%%%%%%%%%%%%%%%%%
% Abstract
%%%%%%%%%%%%%%%%%%%%%%%%%%%%%%%%%%%%%%%%%%%%%%%%%%%%%%%%%%%%%%%%%%%%%%
\clearpage
\begin{abstract}
Analytic energy gradients are presented for a variational two-electron
reduced-density-matrix-driven complete active space self-consistent field
(v2RDM-CASSCF) procedure that employs the density-fitting (DF)
approximation to the two-electron repulsion integrals.  The DF
approximation significantly reduces the computational cost of v2RDM-CASSCF
gradient evaluation, in terms of both the number of floating-point
operations and memory requirements, enabling geometry optimizations on
much larger chemical systems than could previously be considered at the
this level of theory [E. Maradzike {\em et al.}, {\em J. Chem. Theory
Comput.}, 2017, {\bf 13}, 4113--4122].  The efficacy of v2RDM-CASSCF for
computing equilibrium geometries and harmonic vibrational frequencies is
assessed using a set of 25 small closed- and open-shell molecules.
Equilibrium bond lengths from v2RDM-CASSCF differ from those obtained from
configuration-interaction-driven CASSCF (CI-CASSCF) by 0.62 pm and 0.05
pm, depending on whether the optimal reduced-density matrices from
v2RDM-CASSCF satisfy two-particle $N$-representability conditions (PQG) or
PQG plus partial three-particle conditions (PQG+T2), respectively.
Harmonic vibrational frequencies, which are obtained by finite differences
of v2RDM-CASSCF analytic energy gradients, similarly demonstrate that
quantitative agreement between v2RDM- and CI-CASSCF requires the
consideration of partial three-particle $N$-representability conditions.
Lastly, optimized geometries are obtained for the lowest-energy singlet
and triplet states of the linear polyacene series up to dodecacene
(\ce{C50H28}), in which case the active space is comprised of 50 electrons
in 50 orbitals. The v2RDM-CASSCF singlet-triplet energy gap extrapolated
to an infinitely-long linear acene molecule is found to be 7.8 \kcal.

%also evaluated for these molecules, and, as in the case of equilibrium
%bond lengths, the frequencies obtained from computations that enforce the
%PQG+T2 conditions are significantly more accurate than those obtained from
%computations that enforce the PQG conditions alone.  The DF approximation
%significantly reduces the computational cost of v2RDM-CASSCF gradient
%evaluation, in terms of both the number of floating-point operations and
%memory requirements.  As a result, the present implementation is
%applicable to large systems with large active spaces, such as dodecacene
%(\ce{C50H28}) with an active space comprised of 50 electrons in 50
%orbitals.  Optimized geometries are obtained for the lowest-energy singlet
%and triplet states of the linear polyacene series up to dodecacene, 
%and
%the singlet-triplet energy gap extrapolated to an infinite linear chain is
%found to be 7.8 \kcal, at the v2RDM-CASSCF level of theory.

%We demonstrate the accuracy of
%the approach by comparing the v2RDM-CASSCF-optimized geometries and harmonic
%vibrational frequencies for a set of small molecules computed by
%v2RDM-CASSCF to those computed by conventional
%configuration-interaction-driven CASSCF (CI-CASSCF) and those derived from
%experiment.  The mean unsigned differences in bond lengths are 0.62 pm and
%0.05 pm for geometries obtained from v2RDM-CASSCF computations that
%enforce two-particle $N$-representability conditions conditions (PQG) and
%PQG plus partial three-particle conditions (PQG+T2), respectively, when
%compared to those from CI-CASSCF.  

\end{abstract}

\maketitle

%%%%%%%%%%%%%%%%%%%%%%%%%%%%%%%%%%%%%%%%%%%%%%%%%%%%%%%%%%%%%%%%%%%%%%
% Introduction
%%%%%%%%%%%%%%%%%%%%%%%%%%%%%%%%%%%%%%%%%%%%%%%%%%%%%%%%%%%%%%%%%%%%%%
\clearpage
\section{Introduction}
\label{SEC:INTRO}
Nondynamical correlation effects in large molecular systems are
notoriously difficult to model, particularly as the number of
strongly-correlated electrons increases. For small systems, the complete
active space self-consistent field (CASSCF)
method\cite{Roos:1980:157,Siegbahn:1980:323,Siegbahn:1981:2384,Roos:1987:399}
provides a reliable zeroth-order description of the electronic structure
that can be improved by the additional consideration of dynamical
correlation effects, for example, through perturbation theory
\cite{Andersson:1990:5483}.  However, the steep computational scaling of
configuration-interaction (CI) based descriptions of the electronic
structure of the active space limits the applicability of CI-CASSCF to
active spaces comprised of at most 20 electrons in 20
orbitals.\cite{Vogiatzis:2017:184111} As a result, several approximations
to CASSCF that are also based on a CI-type {\em ansatz} have been
proposed, including the restricted active space self-consistent
field,\cite{Olsen:1988:2185,Malmqvist:1990:5477} the generalized active
space (GAS) self-consistent field,\cite{Fleig:2001:4775,Ma:2011:044128}
the split GAS,\cite{Manni:2013:3375} the occupation-restricted multiple
active spaces self-consistent field,\cite{Ivanic:2003:9364} and full CI
quantum Monte Carlo self-consistent field
methods.\cite{Thomas:2015:5316,Manni:2016:1245}  While these methods are
applicable to larger active spaces than are permitted by current CI-CASSCF
implementations, abandoning the CI-based {\em ansatz} altogether allows
one to achieve formally polynomially-scaling approximations to CASSCF.
For example, one of the most popular alternatives to the CI
parameterization of the wave function expansion is the density-matrix
renormalization group (DMRG) approach, wherein the wave function is
expressed as a matrix product
state.\cite{White:1992:2863,White:1993:10345,Schollwock:2005:259,Schollwock:2011:96,Mitrushenkov:2003:4148,Chan:2002:4462,Marti:2008:014104,Zgid:2008:014107,Kurashige:2009:234114,OlivaresAmaya:2015:034102,Szalay:2015:1342,Wouters:2014:272,Knecht:2016:244}
An approximation to CASSCF can be achieved by coupling a DMRG calculation
within an active space to an orbital optimization
scheme.\cite{Ghosh:2008:144117,Yanai:2009:2178,Yanai:2010:024105,Wouters:2014:1501,Sun:2017:291,Ma:2017:2533}
The result is a polynomially-scaling method that can treat large active
spaces required, for example, in extended $\pi$-conjugated molecules,
transition metal dimers, and organometallic
complexes.\cite{OlivaresAmaya:2015:034102}

Alternatively, the electronic structure of the active space can be
described without considering {\em any} wave function parameterization.
The key to this strategy is the realization that the electronic
Hamiltonian contains at most two-body interactions, and, as such, the
electronic energy can be evaluated exactly with knowledge of the
two-electron reduced-density matrix (2RDM).  The allure of replacing the
wave function with the 2RDM lies in the fact that the latter offers a far
more compact representation of electronic structure than that offered by
the exact wave function.  The 2RDM can be determined directly by
minimizing the energy with respect to variations in its elements, subject
to a set of $N$-representability conditions, which are constraints placed
on the 2RDM to ensure that it is derived from an $N$-electron wave
function.\cite{Garrod:1964:1756,Garrod:1975:868,Mihailovic:1975:221,Rosina:1975:300,Erdahl:1979:1366,Erdahl:1979:147,Nakata:2001:8282,Mazziotti:2001:042113,Mazziotti:2002:062511,Mazziotti:2006:032501,Zhao:2004:2095,Fukuda:2007:553,Cances:2006:064101,Verstichel:2009:032508,FossoTande:2016:423,Verstichel:2011:1235}
By coupling an orbital optimization to a variational 2RDM (v2RDM) based
description of the active space, one can achieve a v2RDM-driven
approximation to CASSCF that scales polynomially with respect to the
number of active
orbitals.\cite{Gidofalvi:2008:134108,FossoTande:2016:2260}

We have recently described software to compute v2RDM-CASSCF
energies\cite{FossoTande:2016:2260} and analytic energy
gradients,\cite{Maradzike:2017:4113} which is avaiable as a
plugin\cite{v2rdm:plugin:github} to the \textsc{Psi4} electronic structure
package.\cite{Turney:2012:556,Parrish:2017:3185}  This implementation has
been applied to energy computations involving active spaces as large as 50
electrons in 50 orbitals with the simultaneous optimization of 1892
orbitals.\cite{FossoTande:2016:2260}  The consideration of large numbers
of external orbitals in energy computations is facilitated by the use of
the density-fitting (DF) approximation to the electron repulsion integrals
(ERIs),\cite{Whitten:1973:4496,Dunlap:1979:3396,Feyereisen:1993:359,Vahtras:1993:514}
which leads to significant decreases in the storage requirements and
floating-point cost of the orbital optimization procedure.  However, the
analytic energy gradient implementation described in Ref.
\citenum{Maradzike:2017:4113} employs only conventional ERIs, which limits
its applicability to modestly sized systems and basis sets.  In this
paper, we describe a new implementation of analytic energy gradients for
v2RDM-CASSCF that employs the DF approximation to the ERIs and is thus
applicable to much larger molecular systems.  This new implementation is
available in version 5.1 of the Q-Chem electronic structure
package.\cite{QChem4}

%A drawback to the analytic gradient implementation is that it uses
%conventional ERIs which limits the size of the basis set that can be
%considered in v2RDM-CASSCF optimizations.

This paper is organized as follows.  In Sec. \ref{SEC:THEORY}, we review
the theoretical details of v2RDM-CASSCF, summarize the semidefinite
optimization algorithm used within our software, and present the analytic
gradient expressions.  In Sec. \ref{SEC:RESULTS}, we benchmark the method
by comparing v2RDM-CASSCF equilibrium geometries and harmonic vibrational
frequencies for a set of 25 molecules to those computed with conventional
CASSCF and to those derived from experiment.  We then demonstrate the
applicability of the code to large systems by computing the
singlet-triplet energy gap of the linear acene series up to dodecacene
(\ce{C50H28}) using v2RDM-CASSCF optimized geometries.  Computational
details can be found in Sec. \ref{SEC:DETAILS}.

%%%%%%%%%%%%%%%%%%%%%%%%%%%%%%%%%%%%%%%%%%%%%%%%%%%%%%%%%%%%%%%%%%%%%%
% Theory
%%%%%%%%%%%%%%%%%%%%%%%%%%%%%%%%%%%%%%%%%%%%%%%%%%%%%%%%%%%%%%%%%%%%%%
\section{Theory}
\label{SEC:THEORY}

In this Section, we summarize the theoretical details underlying the
v2RDM-CASSCF energy optimization procedure and v2RDM-CASSCF analytic
gradient evaluation in the case that the ERIs are represented within
the DF approximation. We employ a set of orthonormal molecular orbitals
(MOs) indexed by $p$, $q$, $r$, and $s$ throughout.  The MOs are
partitioned into a set of inactive (doubly occupied) orbitals indexed by
$i$, $j$, $k$, and $l$; a set of active orbitals
indexed by $t$, $u$, $v$, $w$, $x$, and $y$; and a set of external
orbitals.  When spin labels are employed and these labels are not
explicitly specified as  $\alpha$ or $\beta$, general spin labels are
indicated by $\sigma$, $\tau$, $\kappa$, $\lambda$, $\mu$, and $\nu$. The
auxiliary basis functions employed within the DF approximation are indexed
by $P$, $Q$, $R$, and $S$. For the sake of brevity, we will refer to
v2RDM-CASSCF computations performed subject to PQG or PQG+T2 conditions
simply as PQG and PQG+T2.  Table \ref{tab:notation} summarizes this
notation.
\begin{table}[!htpb]
    \caption{Summary of Notation}
    \label{tab:notation}
    \begin{center}
        {\scriptsize
            \begin{tabular}{l l}
                \hline\hline
                Label & Summary \\
                \hline
                $p,q,r,s$ & general molecular orbitals \\
                $i,j,k,l$ & inactive (doubly occupied) orbitals \\
                $t,u,v,w,x,y$ & active orbitals \\
                $P,Q,R,S$ & auxiliary basis functions \\
                $\sigma,\tau,\kappa,\lambda,\mu,\nu$ & spin functions \\
                PQG & v2RDM-CASSCF subject to the PQG conditions \\
                PQG+T2 & v2RDM-CASSCF subject to the PQG+T2 conditions \\
                \hline\hline
            \end{tabular}
        }
    \end{center}

\end{table}

\subsection{CASSCF energy and density}
\label{sec:energy}
The non-relativistic electronic energy for a many-electron system is given
by
\begin{eqnarray}
\label{eqn:energy}
E = \sum_{pq} (T_{pq} + V_{pq})\gamma_{pq} + \sum_{pqrs} (pq|rs) \Gamma_{pqrs},
\end{eqnarray}
where $T_{pq}$ and $V_{pq}$ represent the electron kinetic energy and
electron-nuclear potential energy integrals, respectively, $(pq|rs)$
represents an element of the ERI tensor, and $\gamma_{pq}$ and
$\Gamma_{pqrs}$ represent elements of the spin-free one-electron
reduced-density matrix (1RDM) and the spin-free 2RDM, respectively.  For a
CASSCF wave function, ${\bm \gamma}$ and ${\bm \Gamma}$ exhibit block
structure based on the partitioning of the orbitals.  The non-zero blocks
of ${\bm \gamma}$ are
\begin{equation}
\gamma_{ij} = 2 \delta_{ij},
\end{equation}
and
\begin{equation}
\gamma_{tu} = {}^{1}D^{t_{\alpha}}_{u_{\alpha}} + {}^{1}D^{t_{\beta}}_{u_{\beta}},
\end{equation}
and the elements of the spin blocks that comprise the active-active block
of ${\bm \gamma}$ are defined as
\begin{eqnarray}
{}^{1}D^{t_{\sigma}}_{u_{\sigma}} = \braket{\Psi | \hat{a}^{\dagger}_{t_{\sigma}}\hat{a}_{u_{\sigma}} | \Psi},
\end{eqnarray}
where $\hat{a}^{\dagger}$ and $\hat{a}$ represent creation and
annihilation operators of second quantization, respectively.  The non-zero
elements of ${\bm \Gamma}$ are
\begin{equation}
\Gamma_{ijkl} = 2\delta_{ij}\delta_{kl} - \delta_{il}\delta_{jk},
\end{equation}
\begin{equation}
\Gamma_{ijtu} = \Gamma_{tuij} = \gamma_{tu}\delta_{ij},
\end{equation}
\begin{equation}
\Gamma_{iutj} = \Gamma_{tjiu} = -\frac{1}{2}\gamma_{tu} \delta_{ij},
\end{equation}
and
\begin{equation}
\Gamma_{tuvw} = \frac{1}{2} \left( 
{}^{2}D^{t_{\alpha}v_{\alpha}}_{u_{\alpha}w_{\alpha}} +
{}^{2}D^{t_{\alpha}v_{\beta}}_{u_{\alpha}w_{\beta}} +
{}^{2}D^{t_{\beta}v_{\alpha}}_{u_{\beta}w_{\alpha}} +
{}^{2}D^{t_{\beta}v_{\beta}}_{u_{\beta}w_{\beta}} 
\right),
\end{equation}
where the elements of the active-space spin-blocks are defined by
\begin{eqnarray}
{}^{2}D^{t_{\sigma}v_{\tau}}_{u_{\sigma}w_{\tau}} & = & 
\braket{\Psi | \hat{a}^{\dagger}_{t_{\sigma}}\hat{a}^{\dagger}_{v_{\tau}}\hat{a}_{w_{\tau}}\hat{a}_{u_{\sigma}} | \Psi}.
\end{eqnarray}
Given the block structure of ${\bm \gamma}$ and ${\bm \Gamma}$, Eq.
\ref{eqn:energy} can be reexpressed as 
\begin{eqnarray}
E & = & E_{\rm core} + E_{\rm active}
\end{eqnarray}
where
\begin{equation}
E_{\rm core} = 2 \sum_{i}(T_{ii} + V_{ii}) + \sum_{ij}[ 2 (ii|jj) - (ij|ij) ]\label{ecore},
\end{equation}
and 
\begin{equation}
E_{\rm active} = \sum_{tu} h_{tu} \gamma_{tu} +  \sum_{tuvw} (tu|vw) \Gamma_{tuvw}.\label{eact}
\end{equation}
Here, the one-electron matrix elements, $h_{tu}$, are defined by
\begin{equation}
h_{tu} = T_{tu} + V_{tu} + \sum_{i}[2(ii|tu)-(iu|it)].\label{oei}
\end{equation}
The ERIs entering Eqs. \ref{ecore}-\ref{oei} are computed using the DF
approximation; we have 
\begin{eqnarray}
(pq|rs) & \approx & (pq|rs)_{\rm DF} = \sum_{P} B^{P}_{pq} B^{P}_{rs},
\end{eqnarray} 
and the coefficients, $B^{P}_{pq}$, are determined using the
Coulomb metric as
\begin{eqnarray}
B^{P}_{pq} & = & \sum_{Q} (pq|Q)V^{-1/2}_{QP}
\end{eqnarray}
where
\begin{eqnarray}
V_{PQ} & = & (P|Q) = \int {\rm d}{\bf r}_1 \int {\rm d}{\bf r}_2 \; \chi_P({\bf r}_1)\chi_Q({\bf r}_2) / r_{12},
\end{eqnarray}
and $\chi_P$ and $\chi_Q$ represent auxiliary basis functions.

%Although the inactive orbitals will change during the v2RDM-CASSCF calculation, the contribution of $E_{\rm core}$ to the total energy and gradient is similar in form to standard restricted Hartree--Fock theory for closed-shell molecules.
%The calculation of $E_{\rm active}$ will be performed via a constrained variational optimization of the active space 2RDM which is coupled to an orbital optimization that minimizes $E$ with respect to all orbital parameters. 

\subsection{$N$-representability}

%A 2RDM that is derived from an $N$-electron wave function is said to be $N$-representable.\cite{Coleman:1963:668}
%A v2RDM optimization that exactly enforces $N$-representability is numerically equivalent to full configuration interaction, but such a procedure becomes intractable for all but the smallest of systems represented within small basis sets.
%Therefore, only approximate $N$-representability is enforced in a typical v2RDM optimization.
%In this section we outline the two-body (PQG) and partial three-body (T2) $N$-representablity constraints that are considered in this work.

In v2RDM-CASSCF, the active-space 2RDM is determined by minimizing the
energy with respect to variations in its elements, subject to constraints
intended to guarantee that the 2RDM is derivable from an ensemble of
antisymmetrized $N$-electron wavefunctions (such a 2RDM is said to be
ensemble $N$-representable \cite{Coleman:1963:668}).  Here, we outline
some necessary ensemble $N$-representability conditions.  First, an
ensemble $N$-representable 2RDM is Hermitian
\begin{eqnarray}
{}^{2}D^{t_{\sigma}u_{\tau}}_{v_{\sigma}w_{\tau}} = {}^{2}D^{v_{\sigma}w_{\tau}}_{t_{\sigma}u_{\tau}},
\end{eqnarray}
and it is antisymmetric with respect to the exchange of its indices
\begin{eqnarray}
{}^{2}D^{t_{\sigma}u_{\tau}}_{v_{\sigma}w_{\tau}} =
- {}^{2}D^{u_{\tau}t_{\sigma}}_{v_{\sigma}w_{\tau}} =
- {}^{2}D^{t_{\sigma}u_{\tau}}_{w_{\tau}v_{\sigma}} =
{}^{2}D^{u_{\tau}t_{\sigma}}_{w_{\tau}v_{\sigma}}.
\end{eqnarray}
Second, the 2RDM should map onto the 1RDM through a set of contractions
given by
\begin{equation}
\label{EQN:contract1}
(N_{\sigma} - 1) {}^{1}D^{t_{\sigma}}_{u_{\sigma}} = \sum_{v} {}^{2}D^{t_{\sigma}v_{\sigma}}_{u_{\sigma}v_{\sigma}},
\end{equation}
and
\begin{equation}
\label{EQN:contract2}
N_{\sigma}{}^{1}D^{t_{\tau}}_{u_{\tau}} = \sum_{v} {}^{2}D^{t_{\tau}v_{\sigma}}_{u_{\tau}v_{\sigma}} \;\;\; {\rm for}\;\;\sigma \neq \tau
\end{equation}
where $N_{\sigma}$ is the number of active electrons with spin $\sigma$.
The trace of the 2RDM must also preserve the number of pairs of electrons,
according to
\begin{equation}
\label{EQN:trace1}
\sum_{tu} {}^{2}D^{t_{\sigma}u_{\sigma}}_{t_{\sigma}u_{\sigma}} = N_{\sigma} (N_{\sigma} - 1),
\end{equation}
and
\begin{equation}
\label{EQN:trace2}
\sum_{tu} {}^{2}D^{t_{\sigma}u_{\tau}}_{t_{\sigma}u_{\tau}} = N_{\sigma}N_{\tau}\;\;\; {\rm for}\;\;\sigma \neq \tau.
\end{equation}
%Note that the constraints represented by Eqs. \ref{EQN:trace1} and
%\ref{EQN:trace2} assume a non-relativistic Hamiltonian.  Should the
%Hamiltonian include spin-orbit coupling, then the trace constraints on the
%individual spin-blocks of the 2RDM would reduce to a single constraint on
%the trace of the total 2RDM.  
Moreover, because the eigenvalues of the 1RDM and 2RDM can be interpreted
as occupation numbers for natural orbitals and geminals, respectively,
both of these matrices should be positive semidefinite:
\begin{eqnarray}
{}^{1}{\bf D} & \succeq &  0 \\
{}^{2}{\bf D} & \succeq & 0
\end{eqnarray}

Additional ensemble $N$-representability conditions can be obtained by
considering the positivity of other reduced-density matrices (RDMs) that
are related to the 1RDM and 2RDM.  For example, the algebra of the
creation and annihilation operators implies linear relations that map the
1RDM and 2RDM to the one-hole RDM (${}^{1}{\bf Q}$), the two-hole RDM
(${}^{2}{\bf Q}$), and the electron-hole RDM (${}^{2}{\bf G}$), the
elements of which are defined as
\begin{equation}
{}^{1}Q^{t_{\sigma}}_{u_{\sigma}} = \braket{\Psi | \hat{a}_{t_{\sigma}} \hat{a}^{\dagger}_{u_{\sigma}} | \Psi},
\end{equation}
\begin{equation}
{}^{2}Q^{t_{\sigma}u_{\tau}}_{v_{\sigma}w_{\tau}} = \braket{\Psi | \hat{a}_{t_{\sigma}} \hat{a}_{u_{\tau}} \hat{a}^{\dagger}_{w_{\tau}} \hat{a}^{\dagger}_{v_{\sigma}}| \Psi},
\end{equation}
and
\begin{equation}
{}^{2}G^{t_{\sigma}u_{\tau}}_{v_{\kappa}w_{\lambda}} = \braket{\Psi | \hat{a}^{\dagger}_{t_{\sigma}} \hat{a}_{u_{\tau}} \hat{a}^{\dagger}_{w_{\lambda}} \hat{a}_{v_{\kappa}}| \Psi}.
\end{equation}
The $N$-representability of the 2RDM requires that each of these matrices
be positive semidefinite; these constraints comprise the ``PQG''
constraints of Garrod and Percus.\cite{Garrod:1964:1756}.  In this work,
we also consider the partial three-body constraint that enforces the
nonnegativity of the three-body matrix, ${\bf
T2}$,\cite{Zhao:2004:2095,Erdahl:1978:697} with elements
\begin{eqnarray}
T2^{t_{\sigma}u_{\tau}v_{\kappa}}_{w_{\lambda}x_{\mu}y_{\nu}} & = & \braket{\Psi |\hat{a}^{\dagger}_{t_{\sigma}}\hat{a}^{\dagger}_{u_{\tau}}\hat{a}_{v_{\kappa}}\hat{a}^{\dagger}_{y_{\nu}}\hat{a}_{x_{\mu}}\hat{a}_{w_{\lambda}} | \Psi} \nonumber \\
 & + & \braket{\Psi |\hat{a}^{\dagger}_{y_{\nu}}\hat{a}_{x_{\mu}}\hat{a}_{w_{\lambda}}\hat{a}^{\dagger}_{t_{\sigma}}\hat{a}^{\dagger}_{u_{\tau}}\hat{a}_{v_{\kappa}} | \Psi} .
\end{eqnarray}
This condition is implied by the nonnegativity of two three-body RDMs (the
two-pariticle-one-hole and one-particle-two-hole RDMs) and is a weaker
constraint than the nonnegativity of the three-body RDMs themselves.
However, ${\bf T2}$ maps onto only 1- and 2-body RDMs, and, as a result,
maintaining its positivity is much less computationally demanding than
enforcing that of three-body RDMs.  A similar constraint (on the ``${\bf
T1}$'' matrix) is implied by the nonnegativity of the three-particle and
three-hole RDMs, but this condition is much weaker than the constraint on
${\bf T2}$ and is thus not considered here.  For a non-relativistic
Hamiltonian, we may exploit the spin-block structure of each of these
RDMs; this structure is described in Refs.~\citenum{FossoTande:2016:423}
and \citenum{FossoTande:2016:2260}.

%The spin structure of these matrices can be found in
%Ref.~\citenum{FossoTande:2016:423}.  
%The spin structure of ${\bf T2}$ can be found in
%Ref.~\citenum{FossoTande:2016:2260}

For a non-relativistic Hamiltonian, the 2RDM can also be constrained to
satisfy ensemble spin-state conditions.  For example, when Eqs.
\ref{EQN:contract1}-\ref{EQN:trace1} are satisfied, we have $\langle
\hat{S}_z \rangle = \frac{1}{2} (N_\alpha - N_\beta)$.  By considering the
expectation value of $\hat{S}^{2}$, we arrive at the equality
\begin{eqnarray}
\label{EQN:s2}
\sum_{tu} {}^{2}D^{t_{\alpha}u_{\beta}}_{t_{\alpha}u_{\beta}} = \frac{1}{2}(N_{\alpha} + N_{\beta}) + \frac{1}{4}(N_{\alpha} - N_{\beta})^{2} - S(S+1),
%\sum_{tu} {}^{2}D^{t_{\sigma}u_{\tau}}_{t_{\sigma}u_{\tau}} = \frac{1}{2}(N_{\sigma} + N_{\tau}) + \frac{1}{4}(N_{\sigma} - N_{\tau})^{2} - S(S+1) \;\;\;{\rm for}\;\; \sigma \neq \tau.
\end{eqnarray}
where $S$ represents the total spin angular momentum quantum number.
Slightly stronger \cite{VanAggelen:2012:014110} spin constraints can be
derived for the case that the state in question is the maximal spin 
projection, $|\Psi\rangle = |\Psi_{S,M_S=S}\rangle$, by considering 
action of the raising operator, $\hat{S}^{+}$, on the wavefunction
\begin{equation}
\label{EQN:maxspin}
\hat{S}^{+}\ket{\Psi_{S,M_S=S}} = 0.
\end{equation}
Equation \ref{EQN:maxspin} implies two sets of constraints given by
\begin{equation}
\langle \Psi_{S,M_S=S} | \hat{a}^\dagger_t \hat{a}_u \hat{S}^{+}|\Psi_{S,M_S=S} \rangle = 0 \;\;\; \forall t,u,
\end{equation}
and
\begin{equation}
\langle \Psi_{S,M_S=S} | \hat{S}^{+}\hat{a}^\dagger_t \hat{a}_u |\Psi_{S,M_S=S} \rangle = 0 \;\;\; \forall t,u 
\end{equation}
which are expressible in terms of the elements of the particle-hole RDM as
\begin{equation}
\sum_{v} {}^{2}G^{v_{\beta}v_{\alpha}}_{t_{\beta}u_{\alpha}} = 0 \;\;\; \forall t,u 
\end{equation}
\begin{equation}
\sum_{v} {}^{2}G^{t_{\beta}u_{\alpha}}_{v_{\beta}v_{\alpha}} = 0 \;\;\; \forall t,u 
\end{equation}
Optimizations performed under these maximal spin constraints yield
essentially the same results as those performed by enforcing the Eq.
\ref{EQN:s2} alone (for the maximal spin state).  We include them
nonetheless because we have found that their presence sometimes improves
the convergence properties of the v2RDM-CASSCF optimizations on open-shell
systems.

\subsection{Semidefinite optimization}
\label{subsec:sdp}
The minimization of Eq. \ref{eact} with respect to the elements of
the active-space 2RDM, subject to the constraints outlined above, is a
semidefinite optimization problem.  The primal formulation of this problem
is
\begin{eqnarray}
\label{eqn:primal}
\text{minimize } & E_{\rm primal} &= {\bf c^T} \cdot {\bf x}, \\\notag
\text{such that} & {\bf A} {\bf x} &= {\bf b}, \\\notag 
\text{and} & M({\bf x}) & \succeq 0
\end{eqnarray}
where the vector ${\bf x}$ is the primal solution vector and the vector
${\bf c}$ contains the one- and two-electron integrals.  The constraint
matrix ${\bf A}$ and constraint vector ${\bf b}$ encode the
$N$-representability conditions that ${\bf x}$ must satisfy.  The mapping
$M({\bf x})$ maps the primal solution onto the set of positive
semidefinite active-space RDMs
\begin{equation}
\label{eqn:vectormatrixmap}
M({\bf x})=\left(\begin{array}{cccccc}
                           {}^1{\bf D}&0&0&0&0&0\\
                           0&{}^1{\bf Q}&0&0&0&0\\
                           0&0&{}^2{\bf D}&0&0&0\\
                           0&0&0&{}^2{\bf Q}&0&0\\
                           0&0&0&0&{}^2{\bf G}&0\\
                           0&0&0&0&0&{\bf T2}\\
\end{array}\right).
\end{equation}
The corresponding dual formulation of the problem is
\begin{eqnarray}
\label{EQN:DUAL}
\text{maximize   } &E_{\rm dual}&  = {\bf b}^T\cdot {\bf y}, \\\notag
\text{such  that } &{\bf z}&  = {\bf c} - {\bf A}^T {\bf y}, \\\notag
\text{and }        &M({\bf z})& \succeq 0,
\end{eqnarray}
where the vectors ${\bf y}$ and ${\bf z}$ are the dual solution vectors.

The optimial RDMs are determined using a boundary-point semidefinite
optimization
algorithm.\cite{Povh:2006:277,Malick:2009:336,Mazziotti:2011:083001} This
approach maximizes the augmented Lagrangian for the dual problem
\begin{equation}
\label{eqn:lagrangian}
\mathcal{L}_{\rm act} = {\bf b}^T {\bf y} - {\bf x}^T({\bf A}^T {\bf y}-{\bf c}+{\bf z}) - \frac{1}{2\mu} ||{{\bf A}^T {\bf y}-{\bf c}+{\bf z}}||^2
\end{equation}
by the following two-step procedure:
\begin{enumerate}
\item Solve ${\bf AA}^T{\bf y} = {\bf A}({\bf c}-{\bf z}) + \mu({\bf
b}-{\bf A}{\bf x})$ for ${\bf y}$ by conjugate gradient methods.
\item Update ${\bf x}$ and ${\bf z}$ by separating ${\bf U} = M(\mu {\bf
x} +{\bf A}^T{\bf y} - {\bf c})$ into its positive and negative components
(by diagonalization). The updated primal and dual solutions ${\bf x}$ and
${\bf z}$ are given by $M({\bf x}) = {\bf U}(+)/\mu$ and $M({\bf z}) =
-{\bf U}(-)$.
\end{enumerate}
The penalty parameter $\mu$ is dynamically updated during the course of the v2RDM calculation.\cite{Mazziotti:2011:083001}
The v2RDM optimization is considered converged when
\begin{equation}
|| {\bf A x} - {\bf b} || < \epsilon_{\rm error},
\end{equation}
\begin{equation}
|| {{\bf A}^T {\bf y}-{\bf c}+{\bf z}}  || < \epsilon_{\rm error},
\end{equation}
and
\begin{equation}
| E_{\rm primal} - E_{\rm dual} | < \epsilon_{\rm gap},
\end{equation}
for given thresholds $\epsilon_{\rm error}$ and $\epsilon_{\rm gap}$.

\subsection{Orbital optimization}
In v2RDM-CASSCF, the energy is minimized with respect to both the elements
of the RDMs and the orbital parameters.  We employ an algorithm in which
the orbitals are optimized after a preselected number of v2RDM iterations
(steps 1 and 2 in Sec.~\ref{subsec:sdp}) or after the v2RDM optimization
converges.  Because the v2RDM-CASSCF energy is invariant to rotations
among inactive, active, or external orbitals, the energy is optimized with
respect to rotations between inactive and active, inactive and virtual,
and active and virtual orbitals.  The optimization utilizes an exponential
parameterization of the orbital transformation matrix ${\bf U} = e^{\bf
K}$, where the skew-symmetric matrix ${\bf K}$ contains the nonredundant
rotation parameters.  The unique matrix elements of ${\bf K}$ can be
organized into the vector ${\bm \kappa}$, and the energy expression,
truncated at second order in ${\bm \kappa}$, is \begin{equation} E({\bm
\kappa}) = E({\bf 0}) + {\bm \kappa}^{T}{\bf g} + \frac{1}{2} {\bm
\kappa}^{T}{\bf B}{\bm \kappa}.  \end{equation} The energy is minimized
with respect to the orbital parameters using a quasi-Newton approach that
only requires the computation of the orbital gradient (${\bf g}$) and
diagonal elements of the orbital Hessian (${\bf B}$).  For details of the
orbital optimization procedure, the reader is referred to
Ref.~\citenum{FossoTande:2016:2260}.  We consider the orbitals to be
converged when the norm of the orbital gradient falls below the threshold
$\epsilon_{\rm ograd}$ and the energy computed before and after the
orbitals optimization step differs by less than $\epsilon_{\rm oene}$.

\subsection{Analytic gradients}
To facilitate the derivation of the analytic gradients, we define the
Lagrangian
\begin{equation}
\mathcal{L} = E_{\rm core} + \mathcal{L}_{\rm act},
\end{equation}
which is stationary with respect to variations in the active-space 1-RDM
and 2-RDM (the reader is referred to Ref. \citenum{Maradzike:2017:4113}
for a discussion on the stationarity of $\mathcal{L}_{\rm act}$).  The
gradient of the energy with respect to an arbitrary perturbation $\chi$ is
\begin{eqnarray}
\label{eqn:grad}
\frac{{\rm d} E}{{\rm d} \chi} = \frac{\partial \mathcal{L}}{\partial \chi}
&=& \sum_{pq} (T^{\chi}_{pq} + V^{\chi}_{pq})\gamma_{pq}
+ \sum_{pqrs} (pq|rs)^{\chi}_{\rm DF} \Gamma_{pqrs} \nonumber \\
&-& \sum_{pq} X_{pq} S^{\chi}_{pq}
\end{eqnarray}
where $T^{\chi}_{pq}$, $V^{\chi}_{pq}$, and $S^{\chi}_{pq}$ are the
kinetic energy, electron-nucleus potential energy, and overlap derivative
integrals, respectively.  The term involving the electron repulsion
derivative integrals, $(pq|rs)^{\chi}_{\rm DF}$, is evaluated
as\cite{Weigend:1997:331}
\begin{eqnarray}
\label{EQN:dERI}
\sum_{pqrs} (pq|rs)^{\chi}_{\rm DF} \Gamma_{pqrs} & = & 
2 \sum_{pq}\sum_{P} \Gamma^{P}_{pq} (P|pq)^{\chi} \nonumber \\
&-& \sum_{PQ} \Gamma_{PQ}V^{\chi}_{PQ}
\end{eqnarray}
where
\begin{eqnarray}
\Gamma^{P}_{pq} & = & \sum_{rs}\sum_{Q} \Gamma_{pqrs}B^{Q}_{rs}V^{-1/2}_{QP},
\end{eqnarray}
and
\begin{eqnarray}
\Gamma_{PQ} & = & \sum_{pq}\sum_{R} \Gamma^{P}_{pq} \Gamma^{R}_{pq} V^{-1/2}_{RQ}.
\end{eqnarray}

Although we present the gradient expressions in the MO basis, in practice,
the gradient is evaluated in the AO basis.  As such, the 1RDM and 2RDM
must be transformed to the AO basis before contraction with the derivative
integrals.  Within the DF approximation, only two- and three-index
quantities enter Eq. \ref{eqn:grad}, meaning that we avoid the cost
associated with transforming the full 2RDM to the AO basis, as was done in
our previous implementation. This restructuring of the algorithm results
in tremendous computational savings for derivative computations on large
systems.

The last term in Eq. \ref{eqn:grad} arises from the orbital response to the
perturbation.  It can be shown\cite{Rice:1986:963} that for a CASSCF wave
function with an energy that is stationary with respect to rotations
between all nonredundant orbital pairs, the orbital response depends only
on the overlap derivative integrals and the orbital Lagrangian, ${\bf X}$,
with matrix elements \begin{equation} X_{pq} = \sum_{r} (T_{pr} + V_{pr})
\gamma_{rq} + 2 \sum_{rst} (pr|st) \Gamma_{qrst} \end{equation}

%%%%%%%%%%%%%%%%%%%%%%%%%%%%%%%%%%%%%%%%%%%%%%%%%%%%%%%%%%%%%%%%%%%%%%
% Computational Details
%%%%%%%%%%%%%%%%%%%%%%%%%%%%%%%%%%%%%%%%%%%%%%%%%%%%%%%%%%%%%%%%%%%%%%
\section{Computational details}
\label{SEC:DETAILS}
All v2RDM-CASSCF calculations were carried out in development version of
Q-Chem 5.1.  For geometry optimizations, v2RDM-CASSCF calculations were
considered converged when $\epsilon_{\rm error} < 1.0 \times 10^{-6}$,
$\epsilon_{\rm gap} < 1.0 \times 10^{-4}\;E_{\rm h}$, $\epsilon_{\rm
ograd} < 1.0 \times 10^{-6}\;E_{\rm h}$, and $\epsilon_{\rm oene} < 1.0
\times 10^{-10}\;E_{\rm h}$.  Geometry optimizations were considered
converged when the maximum gradient component reached $1.5 \times
10^{-5}\;E_{\rm h} a^{-1}_0$ and either the maximum atomic displacement
was less than $6.0 \times 10^{-5}\; a_{0}$ or the energy change of
successive optimization cycles was less than $1.0 \times 10^{-8}\;E_{\rm
h}$.  Harmonic vibrational frequencies were computed by finite differences
of the analytic energy gradients using a 5-point stencil with a
displacement of 0.005~\AA, and the ``sow/reap'' mode in the \textsc{Psi4}
software package was used to generate symmetry-adapted displacements.  For
the frequency calculations, we tightened the convergence criteria to
$\epsilon_{\rm error} < 1.0 \times 10^{-8}$, $\epsilon_{\rm gap} < 1.0
\times 10^{-8}\;E_{\rm h}$, $\epsilon_{\rm ograd} < 1.0 \times 10^{-8}
\;E_{\rm h}$, and $\epsilon_{\rm oene} < 1.0 \times 10^{-11} \;E_{\rm h}$.
However, we note that we encountered some difficulties in converging some
PQG+T2 computations this tightly. In these cases (which are noted in Table
\ref{tab:freqs}), the convergence criteria were loosened to $\epsilon_{\rm
error} < 1.0 \times 10^{-6}$, $\epsilon_{\rm gap} < 1.0 \times 10^{-6}
\;E_{\rm h}$, $\epsilon_{\rm ograd} < 1.0 \times 10^{-6}\;E_{\rm h}$, and
$\epsilon_{\rm oene} < 1.0 \times 10^{-9}\;E_{\rm h}$.  We used finite
difference frequency calculations to estimate the error introduced by the
loose convergence thresholds.  Computations were performed at the PQG
level of theory using both sets of thresholds, and we estimate this error
to be less than \cm, except in the case of the low-frequency mode for HNC
which exhibited larger errors (see Supporting Information).

\begin{table}[!htpb]

    \caption{Term symbols and active spaces for the small molecules that
    comprise our test set.}

    \label{tab:molecules}

    \begin{center}
        {\scriptsize
            \begin{tabular}{l c c }
                \hline\hline
                Molecule & Term & Active Space \\
                \hline
                 BF         & ${}^1\Sigma^{+}$      & (10,8) \\
                 BH         & ${}^1\Sigma^{+}$      & (4,5) \\
                 C$_2$      & ${}^1\Sigma^{+}_{g}$  & (8,8) \\
                 CH$_2$         & ${}^1A_{1}$           & (6,6) \\
                 CH$_4$         & ${}^1A_{1}$           & (8,8) \\
                 CO         & ${}^1\Sigma^{+}$      & (10,8) \\
                 F$_2$      & ${}^1\Sigma^{+}_{g}$  & (14,8) \\
                 H$_2$O         & ${}^1A_{1}$           & (8,6) \\
                 HCN            & ${}^1\Sigma^{+}$      & (10,9) \\
                 HF         & ${}^1\Sigma^{+}$      & (8,5) \\
                 HNC            & ${}^1\Sigma^{+}$      & (10,9) \\
                 HNO            & ${}^1A^{\prime}$      & (12,9) \\
                 HOF            & ${}^1A^{\prime}$      & (14,9) \\
                 N$_2$      & ${}^1\Sigma^{+}_{g}$  & (10,8) \\
                 N$_2$H$_2$  & ${}^1A_{g}$          & (12,10) \\
                 NH$_3$         & ${}^1A_{1}$           & (8,7) \\
                 BO         & ${}^2\Sigma^{+}$      & (9,8) \\
                 CH         & ${}^2\Pi$             & (5,5) \\
                 NH$_2$         & ${}^2B_{1}$           & (7,6) \\
                 OH         & ${}^2\Pi$             & (7,5) \\
                 B$_2$      & ${}^3\Sigma^{-}_{g}$  & (6,8) \\
                 CH$_2$         & ${}^3B_{1}$           & (6,6) \\
                 NF         & ${}^3\Sigma^{-}$      & (12,8) \\
                 NH         & ${}^3\Sigma^{-}$      & (6,5) \\
                 O$_2$      & ${}^3\Sigma^{-}_{g}$  & (12,8) \\
                \hline\hline
            \end{tabular}
        }
    \end{center}
\end{table}

All CI-CASSCF calculations were performed using the GAMESS software
package.\cite{Schmidt:1993:1347} The CI-CASSCF calculations were
considered converged when the maximum asymmetry in the Lagrangian matrix
fell below $1.0 \times 10^{-7}\;E_{\rm h}$ and the energy change was
smaller than $1.0 \times 10^{-10}\;E_{\rm h}$.  The CI-CASSCF geometry
optimizations were considered converged when the largest component of the
gradient was below $1.0 \times 10^{-7}\;E_{\rm h} a^{-1}_0$ and the root
mean square gradient was less than $\frac{1}{3} \times 10^{-7}\;E_{\rm h}
a^{-1}_0$.  Harmonic vibrational frequencies were computed with GAMESS
using analytic hessians, which are available for basis sets comprised of
$s$, $p$, and $d$ functions.  Therefore, we report harmonic frequencies
for the cc-pVDZ basis set only.

All calcualtions employed the cc-pVXZ\cite{Dunning:1989:1007} (X = D, T,
Q) basis sets. The cc-pVXZ-JK\cite{Weigend:2002:4285} auxiliary basis sets
were used in the DF approximation for the v2RDM-CASSCF computations.  The
cc-pVDZ-JK basis set is formed by removing the highest angular momentum
functions from the cc-pVTZ-JK basis set.

%%%%%%%%%%%%%%%%%%%%%%%%%%%%%%%%%%%%%%%%%%%%%%%%%%%%%%%%%%%%%%%%%%%%%%
% Results and Discussion
%%%%%%%%%%%%%%%%%%%%%%%%%%%%%%%%%%%%%%%%%%%%%%%%%%%%%%%%%%%%%%%%%%%%%%
\section{Results and discussion}
\label{SEC:RESULTS}

\subsection{Benchmark computations: equilibrium geometries}

We optimized the geometries for the 25 molecules listed in Table
\ref{tab:molecules} using CI- and v2RDM-CASSCF with a full-valence active
space.  Table \ref{tab:bond_length_errors} presents the error in the CI-
and v2RDM-CASSCF bond lengths relative to those derived from experiment.
Agreement with experimental geometries generally improves with the size of
the basis set for both CI- and v2RDM-CASSCF.  For the cc-pVQZ basis set,
the mean unsigned error for the bond lengths are 1.0 pm, 1.6 pm, and 1.1
pm for CI-CASSCF, PQG, and PQG+T2, respectively.  The unsigned errors are
under 2.0 pm for CI-CASSCF and PQG+T2 using the cc-pVQZ basis except for
two cases (\ce{F2} and \ce{B2}).  In general, bond lengths obtained from PQG tend
to deviate more from experiment than those from PQG+T2 or CI-CASSCF.

\begin{table*}[!htpb]

    \caption{Errors in computed equilibrium bond lengths ($\Delta \req$,
    pm)$^{a}$ from CI- and v2RDM-CASSCF using the cc-pVXZ (X = D, T, Q)
    basis sets.  Computed bond lengths are compared to those obtained from
    experiment ($\req$, \AA). All values of \req were taken from
    Ref.~\citenum{CCCBDB} and the references therein.}

    \label{tab:bond_length_errors}

    \begin{center}
        {\scriptsize
            \begin{tabular}{l c c ............. }
                \hline\hline
                & &\multicolumn{12}{c}{$\Delta \req$} \\
                \cline{5-15}
                & & & &\multicolumn{3}{c}{cc-pVDZ} &~& \multicolumn{3}{c}{cc-pVTZ} &~& \multicolumn{3}{c}{cc-pVQZ} &  \\
                \cline{5-7} \cline{9-11} \cline{13-15}
                Molecule & Term & Bond & \req & \multicolumn{1}{c}{CI} & \multicolumn{1}{c}{PQG} &\multicolumn{1}{c}{PQG+T2} &~& \multicolumn{1}{c}{CI} & \multicolumn{1}{c}{PQG} &\multicolumn{1}{c}{PQG+T2} &~& \multicolumn{1}{c}{CI} & \multicolumn{1}{c}{PQG} &\multicolumn{1}{c}{PQG+T2}   \\
                \hline
                 BF         & ${}^1\Sigma^{+}$      & B-F & 1.267 & 2.4  & 2.5 & 2.5   && 0.1  & 0.1  & 0.1  && -0.2 & -0.2 & -0.2 \\
                 BH         & ${}^1\Sigma^{+}$      & H-B & 1.232 & 3.5  & 4.1 & 3.5   && 1.9  & 2.5  & 1.9  && 1.7  & 2.3  & 1.7 \\
                 C$_2$      & ${}^1\Sigma^{+}_{g}$  & C-C & 1.242 & 2.4  & 7.1 & 2.8   && 1.3  & 5.8  & 1.7  && 1.1  & 5.5  & 1.5 \\
                 CH$_2$     & ${}^1A_{1}$           & H-C & 1.107 & 3.2  & 3.6 & 3.2   && 1.8  & 2.3  & 1.8  && 1.7  & 2.2  & 1.7 \\
                 CH$_4$     & ${}^1A_{1}$           & H-C & 1.087 & 2.5  & 3.2 & 2.5   && 1.4  & 2.2  & 1.5  && 1.4  & 2.2  & 1.4 \\
                 CO         & ${}^1\Sigma^{+}$      & C-O & 1.128 & 1.4  & 1.8 & 1.4   && 0.7  & 1.1  & 0.8  && 0.5  & 0.9  & 0.5 \\
                 F$_2$          & ${}^1\Sigma^{+}_{g}$  & F-F & 1.412 & 10.5 & 10.5 & 10.5 && 4.9  & 4.9  & 4.9  && 4.8  & 4.8  & 4.8 \\
                 H$_2$O     & ${}^1A_{1}$           & H-O & 0.958 & 1.3  & 1.4 & 1.3   && 0.6  & 0.7  & 0.6  && 0.5  & 0.6  & 0.5 \\
                 HCN        & ${}^1\Sigma^{+}$      & H-C & 1.064 & 2.4  & 3.0 & 2.4   && -0.7 & 1.9  & -0.7 && -0.7 & -0.6 & -0.7 \\
                            &                       & C-N & 1.156 & 1.9  & 2.8 & 2.0   && 0.4  & 1.7  & 0.5  && 0.3  & 0.8  & 0.3 \\
                 HF         & ${}^1\Sigma^{+}$      & H-F & 0.917 & 0.5  & 0.5 & 0.5   && 0.0  & 0.0  & 0.0  && -0.2 & -0.2 & -0.2 \\
                 HNC        & ${}^1\Sigma^{+}$      & H-N & 0.986 & 2.5  & 3.3 & 2.5   && 1.6  & 2.4  & 1.7  && 1.6  & 2.4  & 1.6 \\
                            &                       & C-N & 1.173 & 1.6  & 2.5 & 1.7   && 0.5  & 1.3  & 0.5  && 0.3  & 1.2  & 0.4 \\
                 HNO        & ${}^1A^{\prime}$      & H-N & 1.090 & -0.2 & 0.0 & -0.2  && -1.3 & -1.1 & -1.2 && -1.4 & -1.3 & -1.3 \\
                            &                       & N-O & 1.209 & 0.6  & 1.4 & 0.7   && 0.2  & 1.0  & 0.3  && 0.0  & 0.8  & 0.1 \\
                 HOF        & ${}^1A^{\prime}$      & H-O & 0.960 & 1.9  & 2.2 & 2.0   && 1.2  & 1.4  & 1.2  && 1.0  & 1.3  & 1.1 \\
                            &                       & O-F & 1.442 & 4.9  & 5.4 & 4.9   && 2.1  & 2.6  & 2.1  && 1.9  & 2.4  & 1.9 \\
                 N$_2$          & ${}^1\Sigma^{+}_{g}$  & N-N & 1.098 & 1.9  & 2.3 & 1.9   && 0.8  & 1.3  & 0.9  && 0.6  & 1.1  & 0.7 \\
                 N$_2$H$_2$ & ${}^1A_{g}$           & H-N & 1.028 & 2.5  & 3.3 & 2.6   && 1.5  & 2.3  & 1.6  && 1.4  & 2.2  & 1.4 \\
                             &                      & N-N & 1.252 & 1.3  & 1.3 & 1.4   && 0.7  & 0.7  & 0.8  && 0.5  & 0.5  & 0.6 \\
                 NH$_3$     & ${}^1A_{1}$           & H-N & 1.012 & 2.2  & 2.5 & 2.2   && 1.1  & 1.5  & 1.1  && 0.9  & 1.3  & 1.0 \\
                 BO          & ${}^2\Sigma^{+}$         & B-O & 1.204 & 1.3  & 1.9 & 1.4   && 0.9  & 1.5  & 1.0  && 0.6  & 1.2  & 0.7 \\
                 CH         & ${}^2\Pi$                 & H-C & 1.120 & 2.7  & 2.5 & 2.7   && 1.3  & 1.2  & 1.3  && 1.1  & 1.0  & 1.1 \\
                 NH$_2$     & ${}^2B_1$                 & H-N & 1.024 & 2.3  & 2.6 & 2.3   && 1.3  & 1.6  & 1.3  && 1.2  & 1.4  & 1.2 \\
                 OH         & ${}^2\Pi$                 & H-O & 0.970 & 1.3  & 1.3 & 1.3   && 0.5  & 0.5  & 0.5  && 0.3  & 0.3  & 0.3 \\
                 B$_2$          & ${}^3\Sigma^{-}_{g}$  & B-B & 1.590 & 3.9 & 8.2 & 4.3    && 2.7  & 7.0  & 3.1  && 2.4  & 6.7  & 2.9 \\
                 CH$_2$     & ${}^3B_1$                 & H-C & 1.085 & 1.7  & 1.9 & 1.7   && 0.5  & 0.7  & 0.5  && 0.4  & 0.7  & 0.4 \\
                 NF         & ${}^3\Sigma^{-}$      & N-F & 1.317 & 1.7  & 1.8 & 1.7   && 0.9  & 1.0  & 0.9  && 0.8  & 0.9  & 0.8 \\
                 NH         & ${}^3\Sigma^{-}$      & H-N & 1.036 & 2.3  & 2.3 & 2.3   && 1.1  & 1.1  & 1.1  && 1.0  & 1.0  & 1.0 \\
                 O$_2$          & ${}^3\Sigma^{-}_{g}$  & O-O & 1.208 & 1.3  & 2.0 & 1.4   && 1.0  & 1.7  & 1.1  && 0.8  & 1.4  & 0.9 \\
                \hline
                            & & MSE$^b$ & - & 2.3  & 3.0  & 2.4  && 1.0 & 1.8 & 1.1 && 0.9 & 1.5 & 0.9  \\
                            & & MUE$^c$ & - & 2.3  & 3.0  & 2.4  && 1.2 & 1.8 & 1.2 && 1.0 & 1.6 & 1.1  \\
                            & & Max$^d$ & - & 10.5 & 10.5 & 10.5 && 4.9 & 7.0 & 4.9 && 4.8 & 6.7 & 4.8  \\
                \hline\hline
            \end{tabular}
        }
    \end{center}
    { \scriptsize
        ${}^a$ $\Delta \req = \req^{\rm CASSCF} - \req$. ${}^b$ mean signed error. ${}^c$ mean unsigned error. ${}^d$ maximum unsigned error.
    }
\end{table*}

Table \ref{tab:angle_errors} provides errors in the CI- and v2RDM-CASSCF
bond angles relative to those derived from experiment.  Again, in general,
these errors decrease with the size of the basis set. For the cc-pVQZ
basis set the errors are all below 5.0\degree\ with mean unsigned errors
of 1.5\degree, 1.7\degree, and 1.5\degree\ for CI-CASSCF, PQG, and PQG+T2,
respectively.  For this test set, v2RDM- and CI-CASSCF provide predictions
in bond angles that are generally similar in quality, when comparing to
angles derived from experiment. For example, CI-CASSCF and PQG+T2 both
underestimate all bond angles with the exception of the H--N--O angle in
HNO. The H--N--N angle in N$_2$H$_2$ is also overestimated when using PQG
within the cc-pVTZ and cc-pVQZ basis sets.  Although the maximum error for
each level of theory exceeds 4.0\degree\; (\ce{CH2}), all other bond
angles agree with those from experiment to within 2.0\degree.

\begin{table*}[!htpb]

    \caption{Errors in computed equilibrium bond angles ($\Delta \aeq$,
    degrees)$^a$ from CI- and v2RDM-CASSCF using the cc-pVXZ (X=D,T,Q)
    basis sets.  Computed bond angles are compared to those obtained from
    experiment (\aeq, degrees). All values of \aeq were taken from
    Ref.~\citenum{CCCBDB} and the references therein.}

    \label{tab:angle_errors}

    \begin{center}
        {\scriptsize
            \begin{tabular}{l c c . ............ }
                \hline\hline
                & & &\multicolumn{12}{c}{$\Delta \aeq$} \\
                \cline{5-15}
                & & & &\multicolumn{3}{c}{cc-pVDZ} &~& \multicolumn{3}{c}{cc-pVTZ} &~& \multicolumn{3}{c}{cc-pVQZ} &  \\
                \cline{5-7} \cline{9-11} \cline{13-15}
                Molecule & Term & Bond & \aeq &\multicolumn{1}{c}{CI} &\multicolumn{1}{c}{PQG} &\multicolumn{1}{c}{PQG+T2} & ~&\multicolumn{1}{c}{CI} &\multicolumn{1}{c}{PQG} &\multicolumn{1}{c}{PQG+T2} &~&\multicolumn{1}{c}{CI}&\multicolumn{1}{c}{PQG} &\multicolumn{1}{c}{PQG+T2}   \\
                \hline
                 CH$_2$     & ${}^1A_{1}$       & H-C-H     & 102.4 & -2.4 & -1.8 & -2.4 && -1.4 & -1.0 & -1.4 && -1.2 & -0.9 & -1.2 \\
                 H$_2$O     & ${}^1A_{1}$       & H-O-H     & 104.5 & -3.3 & -3.5 & -3.3 && -1.9 & -2.1 & -1.9 && -1.6 & -1.8 & -1.6 \\
                 HNO        & ${}^1A^{\prime}$  & H-N-O     & 108.0 &  0.5 &  1.1 &  0.5 &&  0.9 &  1.4 &  0.9 &&  1.0 &  1.5 &  1.0 \\
                 HOF        & ${}^1A^{\prime}$  & H-O-F     & 97.2  & -1.3 & -1.6 & -1.3 && -0.2 & -0.5 & -0.2 && -0.1 & -0.4 & -0.1 \\
                 N$_2$H$_2$ & ${}^1A_g$             & H-N-N     & 106.3 & -1.3 & -0.1 & -1.3 && -0.6 &  0.5 & -0.7 && -0.5 &  0.7 & -0.5 \\
                 NH$_3$     & ${}^1A_{1}$       & H-N-H     & 106.7 & -3.9 & -3.6 & -3.9 && -2.1 & -2.1 & -2.1 && -1.8 & -1.7 & -1.8 \\
                 NH$_2$     & ${}^2B_1$                 & H-N-H     & 103.4 & -3.2 & -3.1 & -3.2 && -1.9 & -2.0 & -1.9 && -1.6 & -1.7 & -1.7 \\
                 CH$_2$     & ${}^3B_1$             & H-C-H     & 135.5 & -4.5 & -4.9 & -4.5 && -4.2 & -4.5 & -4.2 && -4.2 & -4.5 & -4.2 \\
                \hline
                            & & MSE$^b$  & - & -2.4 & -2.2 & -2.4 && -1.4 & -1.3 & -1.5 && -1.2 & -1.1 & -1.3  \\
                            & & MUE$^c$  & - &  2.5 &  2.5 &  2.5 &&  1.7 &  1.8 &  1.7 &&  1.5 &  1.7 &  1.5  \\
                            & & Max$^d$  & - &  4.5 &  4.9 &  4.5 &&  4.2 &  4.5 &  4.2 &&  4.2 &  4.5 &  4.2  \\
                \hline\hline
            \end{tabular}
        }
    \end{center}
    { \scriptsize
        ${}^a$ $\Delta \aeq = \aeq^{\rm CASSCF} - \aeq$. ${}^b$ mean signed error. ${}^c$ mean unsigned error. ${}^d$ maximum unsigned error.
    }
\end{table*}

\begin{figure*}[!htpb]
    \begin{center}
	\caption{Difference in equilibrium bond lengths ($\Delta \req$,
	pm; $\Delta \req = \req^{\rm v2RDM} - \req^{\rm CI}$) obtained
	from full-valence v2RDM- and CI-CASSCF using the (a) cc-pVDZ, (b)
	cc-pVTZ, and (c) cc-pVQZ basis sets.  The bond lengths considered
	correspond to those that are provided in Table
	\ref{tab:bond_length_errors}.} \label{fig:bond_lengths}
        \includegraphics{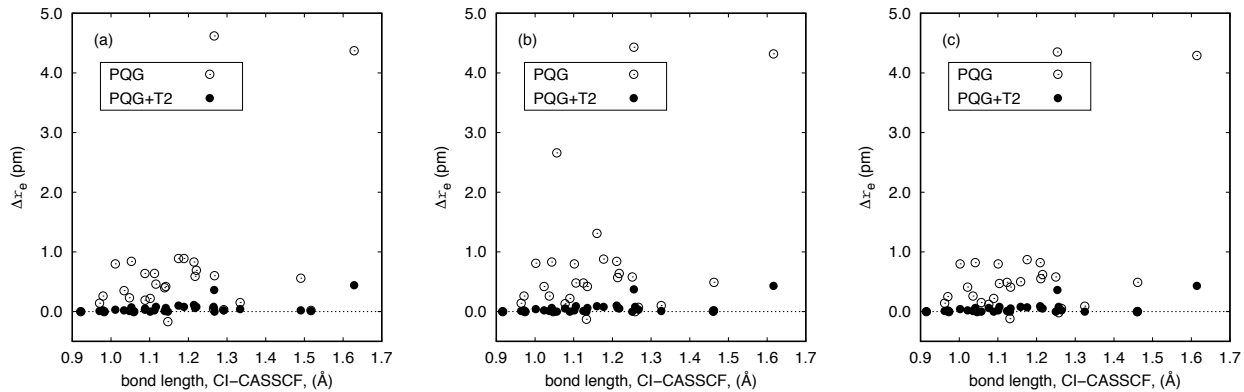}
    \end{center}
\end{figure*}

Figure \ref{fig:bond_lengths} illustrates the difference between the CI-
and v2RDM-CASSCF bond lengths in the cc-pVXZ basis sets (X = D, T, Q).  In
general, these differences are insensitive to the size of the one-electron
basis.  The mean unsigned differences between the CI-CASSCF and PQG bond
lengths are 0.67 pm, 0.74 pm, and 0.62 pm for the cc-pVDZ, cc-pVTZ, and
cc-pVQZ, respectively.  The mean unsigned differences in these basis sets
decrease to, at most, 0.06 pm when the T2 condition is enforced.
Similarly, the mean unsigned difference between the CI- and PQG bond
angles is 0.4\degree\; for all basis sets and falls to 0.0\degree\; when
enforcing the PQG+T2 conditions.  These results demonstrate that the
PQG+T2 conditions lead to quantitative agreement between CI- and
v2RDM-CASSCF geometries.  We recently reported similar deviations between
CI- and v2RDM-CASSCF bond lengths and angles for a test set of 20
molecules with singlet spin states using analytic energy gradients and
conventional (non-DF) ERIs.\cite{Maradzike:2017:4113} The present results
extend these observations to the case of non-singlet molecules and to
gradients computed within the DF approximation.  As seen in
Fig.~\ref{fig:bond_lengths} (and in Ref. \citenum{Maradzike:2017:4113})
v2RDM-CASSCF bond lengths are typically longer than those from CI-CASSCF.
This effect can be rationalized in terms of the ``over correlation''
problem of v2RDM methods; for small molecules, more approximate
$N$-representability conditions lead to longer bond lengths.

%%%%%%%%%%%%%%%%%%%%%%%%%%%%%%%%%%%%%%%%%%%%%%%%%%%%%%%%%%%%%%%%%%%%%%
% v2RDM-CASSCF vs CI-CASSCF Bond Lengths
%%%%%%%%%%%%%%%%%%%%%%%%%%%%%%%%%%%%%%%%%%%%%%%%%%%%%%%%%%%%%%%%%%%%%%

We note two clear outliers in Fig.  \ref{fig:bond_lengths}, which
correspond to bond lengths for \ce{C2} and \ce{B2} optimized under the PQG
conditions.  The CI-CASSCF wave functions for these two molecules exhibit
the most multconfigurational character in the entire set, as measured by
the magnitude of the largest CI-coefficients in the respective expansions
(0.83 for \ce{C2} and 0.88 for \ce{B2}).  For all other molecules
considered herein, the largest CI-coefficient is greater than 0.95, with
the exception of \ce{CH4}, which has a leading CI-coefficient of 0.92.
These outliers suggest that the PQG conditions are insufficient to
correctly describe the static correlation effects in \ce{B2} and \ce{C2}.

\subsection{Benchmark computations: harmonic vibrational frequencies}
%exception of four cases: (1) the $\Sigma_g$ mode of \ce{F2} (-33\%), (2)
%the $3A^{\prime}$ mode of \ce{HOF} (-16\%), (3) the $2A_{1}$ mode of
%\ce{NH3} (29\%), and (4) the $2A_{1}$ mode of triplet \ce{CH2} (17\%).

In this section, we evaluate the quality of v2RDM-CASSCF harmonic
vibrational frequencies computed from finite differences of analytic
energy gradients.  Table \ref{tab:freqs} presents the error in the
harmonic frequencies obtained from CI- and v2RDM-CASSCF within the cc-pVDZ
basis set, as compared to those derived from experiment.  The mean
unsigned errors are 64 \cm, 84 \cm, and 65 \cm at the CI-CASSCF, PQG,
PQG+T2 levels of theory, respectively.  The percent error is less than 9\%
for all CI-CASSCF frequencies, with the exception of four cases: (1) the
$\Sigma_g$ stretch of \ce{F2} (-33\%), (2) the $3A^{\prime}$ bend of
\ce{HOF} (-16\%), (3) the $2A_{1}$ wagging mode of \ce{NH3} (29\%), and
(4) the $2A_{1}$ bend of triplet \ce{CH2} (17\%).  The frequencies from
PQG+T2 agree with those from CI-CASSCF for these four cases.  For the remaining modes, the
PQG+T2 frequencies similarly agree with those from experiment to within
9\%.  On the other hand, frequencies computed with the PQG conditions
alone are less reliable;  eight modes are predicted incorrectly by more
than 10\%, with the worst offender having a -51\% error (the
doubly-degenerate $\Pi$ bend in HNC).

\begin{table*}[!htpb]

    \caption{Errors in computed harmonic vibrational frequencies ($\Delta
    \weq$, cm$^{-1}$)$^a$ computed using CI- and v2RDM-CASSCF for the
    cc-pVDZ basis set. Computed frequencies are compared to those obtained
    from experiment (\weq, cm$^{-1}$). All values of \weq were taken from
    Ref.~\citenum{CCCBDB} and the references therein.}

    \label{tab:freqs}

    \begin{center}
        {\scriptsize
            \begin{tabular}{l c c c . ... }
                \hline\hline
                & & & & &\multicolumn{3}{c}{$\Delta \weq$} \\
                \cline{6-8}
                %& & & & &\multicolumn{3}{c}{cc-pVDZ}   \\
                %\cline{6-8} \\
                Molecule & Term & Mode & Sym & \weq & \multicolumn{1}{c}{CI} & \multicolumn{1}{c}{PQG} &\multicolumn{1}{c}{PQG+T2}   \\
                \hline
                 BF         & ${}^1\Sigma^{+}$      & 1 & $\Sigma$          & 1402 & -67     & -68      & -68^{b} \\
                 BH             & ${}^1\Sigma^{+}$      & 1 & $\Sigma$          & 2367 & -100    & -139     & -99 \\
                 C$_2$      & ${}^1\Sigma^{+}_{g}$  & 1 & $\Sigma_{\rm g}$  & 1855 & -5      & -363     & -31 \\
                 CH$_2$         & ${}^1A_{1}$           & 1 & A$_1$                 & 2806 & -14     & -45      & -16 \\
                            &                       & 2 & A$_1$             & 1353 & 64      & 9        & 64 \\
                            &                       & 3 & B$_2$             & 2865 & -11     & -19      & -12 \\
                 CH$_4$         & ${}^1A_{1}$           & 1 & A$_1$                 & 2917 & 22      & -34      & 20^{b} \\
                            &                       & 2 & E                     & 1534 & 22      & -9       & 22^{b} \\
                            &                       & 3 & T$_2$                 & 3019 & 49      & 66       & 48^{b} \\
                            &                       & 4 & T$_2$                 & 1306 & 34      & 25       & 33^{b} \\
                 CO             & ${}^1\Sigma^{+}$      & 1 & $\Sigma$          & 2170 & -5      & -43      & -11 \\
                 F$_2$      & ${}^1\Sigma^{+}_{g}$  & 1 & $\Sigma_{\rm g}$  & 917  & -304    & -305     & -304 \\
                 H$_2$O         & ${}^1A_{1}$           & 1 & A$_1$                 & 3657 & 65      & 42       & 62 \\
                            &                       & 2 & A$_1$                 & 1595 & 121     & 114      & 121 \\
                            &                       & 3 & B$_2$                 & 3756 & 78      & 55       & 79 \\
                 HCN        & ${}^1\Sigma^{+}$      & 1 & $\Sigma$          & 3312 & 59      & 7        & 55^{b} \\
                             &                          & 2 & $\Sigma$          & 2089 & 6       & -65      & -4^{b} \\
                             &                          & 3 & $\Pi$             & 712  & 10      & -107     & 7^{b} \\
                 HF             & ${}^1\Sigma^{+}$      & 1 & $\Sigma$          & 4138 & -83     & -83      & -83 \\
                 HNC        & ${}^1\Sigma^{+}$      & 1 & $\Sigma$          & 3653 & 84      & -14      & 79^{b} \\
                             &                      & 2 & $\Sigma$          & 2029 & 14      & -59      & 6^{b} \\
                            &                       & 3 & $\Pi$             & 477  & -1      & -245     & -5^{b} \\
                 HNO        & ${}^1A^{\prime}$      & 1 & A$^{\prime}$      & 2684 & -48     & -26      & -54 \\
                             &                      & 2 & A$^{\prime}$      & 1565 & 36      & 15       & 29 \\
                             &                      & 3 & A$^{\prime}$      & 1501 & 39      & 42       & 33 \\
                 HOF        & ${}^1A^{\prime}$      & 1 & A$^{\prime}$      & 3537 & 125     & 88       & 121^{b} \\
                             &                      & 2 & A$^{\prime}$      & 1393 & -67     & -80      & -69^{b} \\
                             &                      & 3 & A$^{\prime}$      & 886  & -140    & -150     & -140^{b} \\
                 N$_2$      & ${}^1\Sigma^{+}_{g}$  & 1 & $\Sigma_{\rm g}$  & 2359 & -4      & -54      & -15 \\
                 N$_2$H$_2$ & ${}^1A_{g}$           & 1 & A$_{\rm g}$       & 3058 & 37      & -65      & 26^{b} \\
                            &                       & 2 & A$_{\rm g}$       & 1583 & 37      & 30       & 35^{b} \\
                            &                       & 3 & A$_{\rm g}$       & 1529 & 20      & 16       & 12^{b} \\
                            &                       & 4 & A$_{\rm u}$       & 1289 & 27      & -10      & 25^{b} \\
                            &                       & 5 & B$_{\rm u}$       & 3120 & 3       & -73      & -5^{b} \\
                            &                       & 6 & B$_{\rm u}$       & 1316 & 37      & 28       & 35^{b} \\
                 NH$_3$     & ${}^1A_{1}$           & 1 & A$_1$             & 3337 & -9      & -52      & -13 \\
                            &                       & 2 & A$_1$             & 950  & 279     & 257      & 280 \\
                            &                       & 3 & E                 & 3444 & 8       & -18      & 5  	\\
                            &                       & 4 & E                 & 1627 & 81      & 62       & 81 	\\
                 BO             & ${}^2\Sigma^{+}$      & 1 & $\Sigma$          & 1886 & -14     & -61      & -19^{b} \\
                 CH             & ${}^2\Pi$                 & 1 & $\Sigma$          & 2859 & -114    & -83      & -114^{b} \\
                 NH$_2$         & ${}^2B_1$                 & 1 & A$_1$                 & 3219 & -5      & -30      & -7^{b} \\
                            &                       & 2 & A$_1$                 & 1497 & 76      & 68       & 76^{b} \\
                            &                       & 3 & B$_2$                 & 3301 & 6       & -19      & 5^{b} \\
                 OH             & ${}^2\Pi$                 & 1 & $\Sigma$          & 3738 & -130    & -130     & -130 \\
                 B$_2$      & ${}^3\Sigma^{-}_{g}$  & 1 & $\Sigma_{\rm g}$  & 1051 & -32     & -173     & -51 \\
                 CH$_2$         & ${}^3B_1$             & 1 & A$_1$             & 2806 & 245     & 218      & 245^{b} \\
                            &                       & 2 & A$_1$             & 963  & 160     & 193      & 160^{b} \\
                            &                       & 3 & B$_2$             & 3190 & 71      & 42       & 70^{b} \\
                 NF             & ${}^3\Sigma^{-}$      & 1 & $\Sigma$          & 1141 & -65     & -67      & -66^{b} \\
                 NH             & ${}^3\Sigma^{-}$      & 1 & $\Sigma$          & 3282 & -171    & -171     & -171 \\
                 O$_2$      & ${}^3\Sigma^{-}_{g}$  & 1 & $\Sigma_{\rm g}$  & 1580 & -51     & -108     & -62^{b} \\
                \hline
                            & & & MSE$^c$ & - & 9   & -31  & 6  \\
                            & & & MUE$^d$ & - & 64  &  84  & 65  \\
                            & & & Max$^e$ & - & 304 & 363  & 304  \\
                \hline\hline
            \end{tabular}
        }
    \end{center}
    { \scriptsize
        ${}^a$ $\Delta \weq = \weq^{\rm CASSCF} - \weq$. ${}^b$ ``loose'' convergence criteria used (Sec.~\ref{SEC:DETAILS}). ${}^c$ mean signed error. ${}^d$ mean unsigned error. ${}^e$ maximum unsigned error.
    }
\end{table*}

% AED: this next bit seems out of place.  also, i don't think anything was noted about F2 in the previous section...

%As mentioned in
%the previous subsection, the largest error in bond lengths in the test set
%for CI- and v2RDM-CASSCF was for \ce{F2} where both levels of theory
%predicted a longer bond length than experiment.  Clearly, the lack of
%dynamical correlation in CASSCF with a full-valence active space is
%particularly troublesome for correctly describing the electronic structure
%of \ce{F2}.  Additionally, the largest error in bond angles in the test
%set for CI- and v2RDM-CASSCF was for \ce{CH2} (${}^3B_1$) where both
%levels of theory predicted bond angles over 4\degree\; smaller than
%experiment.

\begin{figure}[!htpb]
    \begin{center}

	\caption{Difference in equilibrium harmonic vibrational
	frequencies ($\Delta \weq = \weq^{\rm v2RDM} -
	\weq^{\rm CI}$) obtained from full-valence v2RDM- and CI-CASSCF
	using the cc-pVDZ basis sets. The cc-pVDZ-JK auxiliary basis set
	was used in the v2RDM-CASSCF optimizations. The frequencies
	considered correspond to those that are provided in Table
	\ref{tab:freqs}.} \label{fig:freq}

        \includegraphics{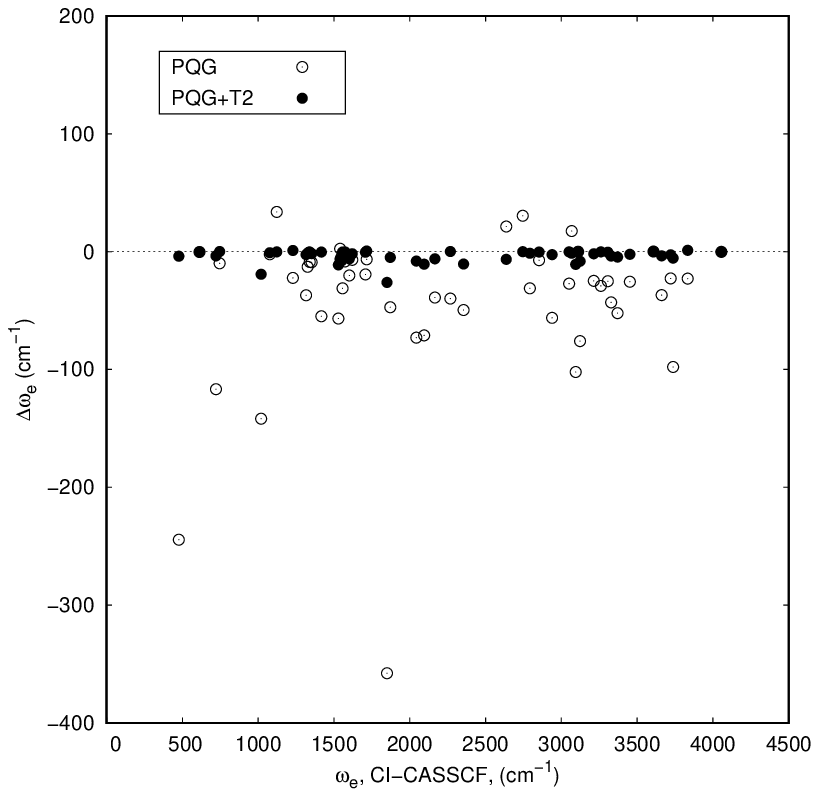}
    \end{center}
\end{figure}

Figure \ref{fig:freq} illustrates the difference between harmonic
frequencies obtained from v2RDM-CASSCF (enforcing both the PQG and PQG+T2
conditions) and CI-CASSCF.  It is clear that the consideration of the T2
condition dramatically improves the agreement between v2RDM- and
CI-CASSCF.  We find only two modes for which the PQG constraints alone
provide better agreement with CI-CASSCF: (1) the $3A^{\prime}$ bend of
\ce{HNO} and (2) the $3A_g$ mode of \ce{N_2H2}.  However, we note that the
discrepancies between CI-CASSCF and PQG+T2 frequencies are quite small in
these cases (less than 10 \cm).  For PQG, the percent difference in the
predicted frequencies, relative to CI-CASSCF, is less than 4\% in all but
four cases: (1) the $\Sigma_g$ stretch of \ce{C2} (-19\%), (2) the $\Pi$
bend of \ce{HCN} (-16\%), (3) the $\Pi$ bend of \ce{HNC} (-51\%), and (4) the $\Sigma_g$ stretch
\ce{B2} (-14\%).  The agreement with CI-CASSCF is significantly improved
upon considering the T2 condition, in which case the percent differences
are less than 1\% in all but two cases: (1) the $\Sigma_g$ stretch of
\ce{C2} (-1\%) and (2)
the $\Sigma_g$ stretch of \ce{B2} (-2\%).  The mean unsigned differences
between v2RDM- and CI-CASSCF frequencies are 44 \cm and 4 \cm when
optimized RDMs satisfy the PQG and PQG+T2 constraints, respectively.  In
general, we note that the v2RDM-CASSCF frequencies are lower than those
predicted by CI-CASSCF.  Of the 52 frequencies considered, only five PQG
and three PQG+T2 frequencies are significantly (more than 0.5 \cm) higher
than the corresponding CI-CASSCF frequency.  It appears that the
over-correlation associated with approximate $N$-representability
manifests itself in generally underestimated harmonic frequencies.

%Based on the results in Fig
%\ref{fig:bond_lengths} and Fig \ref{fig:freq} the ``over correlation''
%problem of v2RDM-CASSCF with approximate $N$-representability conditions
%results in bond lengths that are typically overestimated and frequencies
%that are typically underestimated.

\subsection{Linear acenes: equilibrium geometries and singlet-triplet gap}

The linear polyacene series exhibits complex electronic structure, and an
extensive literature considers the relative ordering of the lowest-energy
singlet and triplet states, as well as the degree to which the singlet
states of larger members of the series can be considered as having
polyradical character.
\cite{Houk:2001:5517,Bendikov:2004:7416,Hachmann:2007:134309,Gidofalvi:2008:134108,Pelzer:2011:5632,Mizukami:2013:401,Plasser:2013:2581,FossoTande:2016:2260,FossoTande:2016:423,Horn:2014:1511,Horn:2015:054302,Yang:2016:E5098,Dupuy:2018:134112}
Here, we demonstrate the applicability of our v2RDM-CASSCF energy gradient
implementation to large molecules with large active spaces by reporting
equilibrium geometries and adiabatic singlet-triplet energy gaps for the
linear polyacene series up to dodecacene.  The active space is chosen to
be comprised of the $\pi$-MO network, which, for an acene molecule
consisting of $k$ fused six-membered rings, corresponds to a ($4k+2$,
$4k+2$) active space.  For dodecacene, the active space consists of 50
electrons in 50 orbitals.

\begin{figure}[!htpb]
    \begin{center}

	\caption{Adiabatic singlet-triplet excitation energies for
	v2RDM-CASSCF PQG, DMRG-CASCI,\cite{Hachmann:2007:134309} and
	JSD\cite{Dupuy:2018:134112} along with vertical excitation
	energies using particle-particle random phase
	approximation\cite{Yang:2016:E5098} and experimental excitation
	energies.\cite{JBBirks:1970,Burgos:1977:249,Sabbatini:1982:3585,Schiedt:1997:201}}

        \label{fig:stgap}
        \includegraphics{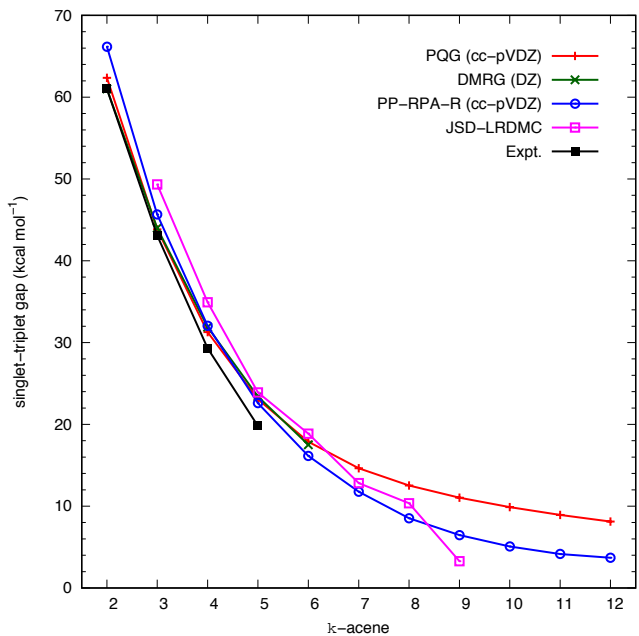} %#[width=3.5in]{stgap.eps}

    \end{center}
\end{figure}

Figure \ref{fig:stgap} shows the adiabatic singlet-triplet excitation
energies computed at the PQG/cc-pVDZ, DMRG-driven complete active space CI
(CASCI)/DZ,\cite{Hachmann:2007:134309} and quantum Monte Carlo (QMC)
\cite{Dupuy:2018:134112} levels of theory, along with vertical excitation
energies derived from the particle-particle random phase approximation
(PP-RPA, using the cc-pVDZ basis)\cite{Yang:2016:E5098} and
experiment.\cite{JBBirks:1970,Burgos:1977:249,Sabbatini:1982:3585,Schiedt:1997:201}
The QMC results taken from Ref. \citenum{Dupuy:2018:134112} were obtained
using a Jastrow single determinant (JSD) wave function, optimized using
lattice regularized diffusion Monte Carlo (LRDMC).  PP-RPA, DMRG-CASCI,
PQG, and experiment are all in reasonable agreement for the smaller
members of the series.  Using the simple exponential decay formula $E = a
e^{-n/b} + c$, the PQG estimate of the gap at the infinitely-long-molecule
limit is 7.8 \kcal.  This value is slightly larger than that recently
estimated\cite{Yang:2016:E5098} using PP-RPA, but direct comparisons to
this and other values is complicated for several reasons.  First, the
PP-RPA excitation energies presented in Ref.  \citenum{Yang:2016:E5098}
are vertical; the singlet and triplet energies reproduced in Fig.
\ref{fig:stgap} were evaluated at the singlet geometry, which was
optimized using restricted B3LYP (hence, the ``R'' in PP-RPA-R).  Second,
comparisons to JSD are difficult because the JSD curve is far from smooth,
and the second kink at nonacene can be attributed to the fact that this
particular geometry was optimized at the JSD level of theory, while all of
the other geometries were optimized using DFT (with the B3LYP
functional).\cite{Dupuy:2018:134112}  Nonetheless, it does appear that the
lack of dynamical correlation effects in our computations may lead to an
overestimation of the singlet-triplet energy gap in the limit of
infinitely long acene molecules.  Indeed, this assertion is consistent
with the recent observation\cite{Mostafanejad:2018:arXiv} that the
singlet-triplet gap closes when v2RDM-CASSCF reference densities are
employed within the multiconfiguration pair-density functional theory
(MCPDFT) approach\cite{Gagliardi2017,LiManni2014}.  The gap at the
infinitely-long-molecule limit presented in Ref.
\citenum{Mostafanejad:2018:arXiv} is $\approx$ 4.8--4.9 \kcal, depending
on the functional employed within MCPDFT.  Reference
\citenum{Mostafanejad:2018:arXiv} is one of several recent
large-active-space MCPDFT-based studies of polyacene
molecules\cite{Ghosh:2017:2741,Sharma:2018:arXiv}.

\begin{figure}[!htpb]
    \begin{center}

    \caption{C--C bond lengths (\AA) along the long edge of the linear
    polyacene molecules. Results are presented for the lowest-energy
    singlet and triplet states computed at the PQG/cc-pVDZ level of
    theory.}

        \label{fig:bla}
        \includegraphics{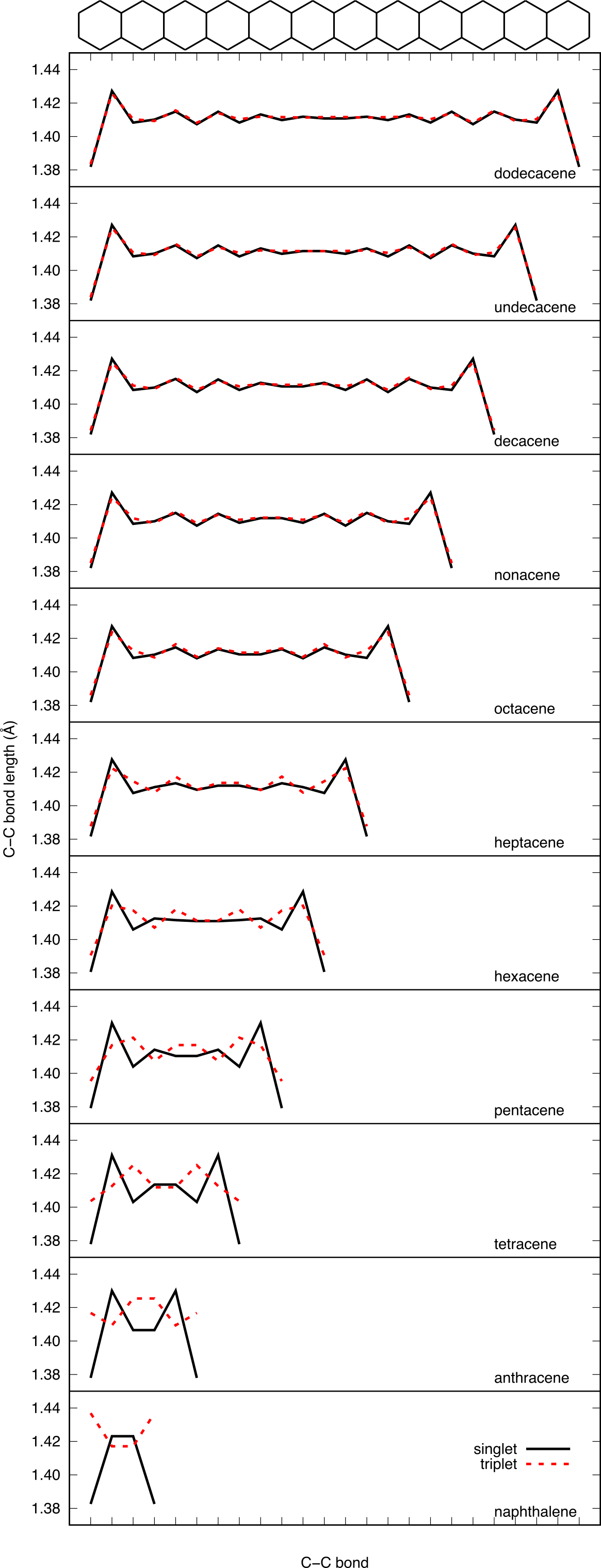}
    \end{center}
\end{figure}

Figure \ref{fig:bla} shows the bond length alternation (BLA) for the C--C
bonds along the long edge of each member of the linear polyacene series.
The equilibrium cartesian coordinates are provided in the
Supporting Information.  In general, the alternation appears greatest
toward the outer parts of the molecules, with smaller changes in the
middle of the molecules; this trend applies to both the singlet and triplet
states and is in agreement with previous work. 
For dodecacene, the C--C bond lengths lie within the range of
1.38 \AA\; to 1.43 \AA\; and approach a bond length of 1.41 \AA\; in the
center of the molecule.  This bond equalization, which has been reported
previously,\cite{Jiang:2008:332,Qu:2009:7909} is more noticeable in the
geometry for the triplet state.  Dupuis {\em et
al.}\cite{Dupuy:2018:134112} note that the equalization is associated with
the localization of charge along the edges of the acene molecule, which,
for the singlet state, is a signature of an antiferromagnetic arrangment
of electrons that could be described as a di- or even polyradical.  The
central C--C bond length limit of 1.41 \AA\; predicted by PQG is
consistent with the limit of 1.406 \AA\; separately reported using a
spin-polarized DFT\cite{Jiang:2008:332} and a DMRG valence bond
model.\cite{Qu:2009:7909}

For the smaller members of the series, the BLA pattern is qualitatively
different for the singlet and triplet states, but, as the length of the
molecules increases, the singlet and triplet BLA patterns become more
similar.  These results constrast with the JSD-derived geometry for
nonacene presented in Ref.  \citenum{Dupuy:2018:134112}. For JSD, the BLA
pattern for the singlet and triplet states of nonacene clearly differ, and
large oscillations (on the order of 0.02 \AA\; for the triplet) persist in
the center of the molecule.  Since the JSD wave function describes
dynamical correlation effects not captured by v2RDM-CASSCF, it is tempting
to attribute these differences to a lack of dynamical correlation in the
present computations.  However, we observe similar agreement between the
BLA patterns of the singlet and triplet states as described by
unrestricted B3LYP, which {\em only} captures dynamical correlation
effects.  Rather, the discrepancies between JSD and PQG can be traced to a
lack of static correlation effects in the former method.  This hypothesis
is partially confirmed by the Jastrow double determinant (JDD) description
of the geometry of the singlet state of nonacene, also provided in Ref.
\citenum{Dupuy:2018:134112}.  JDD captures some static correlation effects
that are missing in JSD, and differences between the predictions of JSD
and JDD can be used to quantify whether or not the longer members of the
acene series have open-shell singlet character. For the singlet state, JDD
predicts that alternations in bond lengths are most apparent at the edge
of the molecule, in agreement with the present PQG results.  However, Ref.
\citenum{Dupuy:2018:134112} also provides results using the Jastrow
antisymmetric geminal power (JAGP) wave function that are more in line
with those of JSD.  The JAGP {\em ansatz} captures captures static
correlation effects beyond those described by JDD, and, apparently, these
additional considerations lead to a reduction in the open-shell character
of the singlet state, as well as a commensurate reduction in the charge
localization along the edges of the acene molecules.  These observations,
which are outlined in Ref. \citenum{Dupuy:2018:134112}, are consistent
with those of Ref. \citenum{Lee:2017:602}, which employed a
coupled-cluster valence-bond singles and doubles (CCVB-SD) description of
the valence space.  At the CCVB-SD level of theory, the polyradical nature
of the longer members of the linear polyacene series is significantly
reduced when the $\sigma$-network is correlated alongside the
$\pi$-network, as compared to the case where the $\pi$-network alone is
correlated.  Because our active space consists of only the $\pi$-network,
signatures of polyradical character, such as the equalization of the BLA
reported here and the natural orbital occupation numbers reported
elsewhere,\cite{FossoTande:2016:423,FossoTande:2016:2260} may be
exaggerated.

%does not include a JDD-derived structure for
%the triplet state of nonacene, so we cannot assess whether or not the BLA
%patterns for singlet and triplet differ at the JDD level of theory.

\begin{figure}[!htpb]
    \begin{center}
    \caption{Root mean square (RMS) difference between the C--C bond
    lengths along the long edge of the linear polyacene molecules,
    as optimized for the lowest-energy singlet and triplet states. }
        \label{fig:bla_rms}
        \includegraphics[width=3.5in]{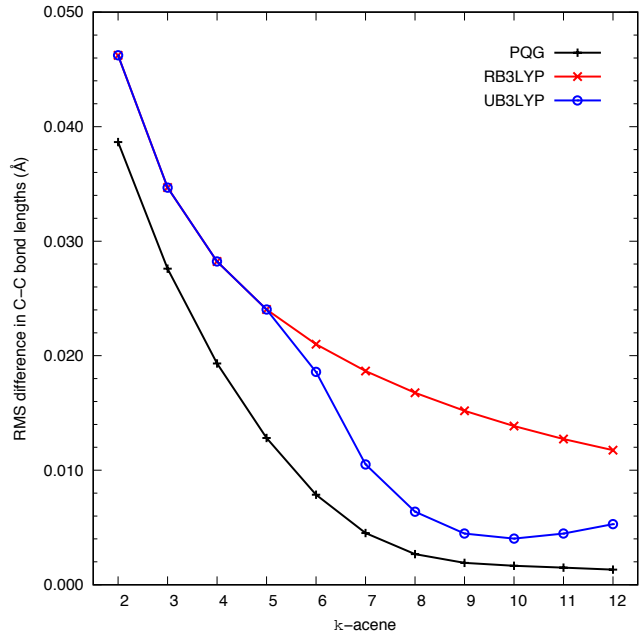}
    \end{center}
\end{figure}

We can further quantify the similarities between the singlet and triplet
structures by comparing the root mean square (RMS) difference in the C--C
bond lengths along the long edge of each linear acene molecule.  Figure
\ref{fig:bla_rms} illustrates this quantity using structures obtained at
the PQG/cc-pVDZ and B3LYP/cc-pVDZ (restricted and unrestricted) levels of
theory. 
%; the B3LYP structures were taken from Ref. \citenum{Yang:2016:E5098}.
%;
%geometries were optimized using each method within the cc-pVDZ basis set.
The RMS difference in the bond lengths decreases monotonically for both
restricted B3LYP and PQG, with the PQG value falling to less than 0.002
\AA\; at dodecacene.  On the other hand, from this metric, it appears that
the singlet and triplet states of the longer members of the series are
predicted to have quite different geometries at the restricted B3LYP level
of theory.  Indeed, when we look at the BLA patterns for restricted B3LYP,
we find that this method recovers some of the characteristics of the
nonacene geometry from JSD: (i) large changes in the bond lengths persist
into the middle of the molecules, and (ii) the BLA patterns for the
singlet and triplet states are qualitatively different.  For molecules as
large or larger than hexacene, spin-broken B3LYP solutions for the singlet
states becomes energetically favorable, which leads to significant
decreases in the RMS difference in the bond lengths.

%in the supporting information and focus on dodecacene here
%as an example.  Figure \ref{fig:bla} shows the bond length alternation
%(BLA) for the C-atoms along one edge of dodecacene.  The C-C bond lengths
%lie within the range of 1.38 \AA\; to 1.43 \AA\; and approach a bond
%length of 1.41 \AA in the center of the molecule.  This bond equalization
%is more noticeable in the optimized geometries for the triplet state.  The
%reduction in BLA is an indicator of diradical/polyradical
%states.\cite{Jiang:2008:332,Qu:2009:7909,Dupuy:2018:134112}

%%%%%%%%%%%%%%%%%%%%%%%%%%%%%%%%%%%%%%%%%%%%%%%%%%%%%%%%%%%%%%%%%%%%%%
% Conclusions
%%%%%%%%%%%%%%%%%%%%%%%%%%%%%%%%%%%%%%%%%%%%%%%%%%%%%%%%%%%%%%%%%%%%%%
\section{Conclusions}
We have presented an implementation of analytic energy gradients for the
v2RDM-driven CASSCF method using the density-fitting approximation to the
electron repulsion integrals.  Benchmark calculations for equilibrium
geometries and harmonic vibrational frequencies indicate that v2RDM-CASSCF
performs as well as CI-CASSCF in reproducing experimental results for a
test set of small molecules.  When the two-particle $N$-representability
conditions are enforced, geometries and frequencies generally agree with
those from CI-CASSCF, and significantly improved results are obtained when
also enforcing three-particle $N$-representability conditions (T2).  In
the current implementation, enforcing the PQG and T2 conditions requires
$\mathcal{O}(n^{6})$ and $\mathcal{O}(n^{9})$, floating-point operations,
respectively, where $n$ is the number of active space orbitals.  For
large-scale applications, the PQG conditions may be the only practically
enforceable ones; fortunately, the present benchmark calculations indicate
that these conditions should be adequate for equilibrium geometries.  Some
care should be taken, however, should one wish to compute harmonic
frequencies under the PQG conditions alone.

The DF approximation facilitates the evaluation of v2RDM-CASSCF analytic
energy gradients for large molecules with active spaces that are much
larger than those that existing CI-CASSCF implementations can reasonably
consider.  We demonstrated this capability by optimizing the geometries
for the lowest-energy singlet and triplet states of the linear polyacene
series up to dodecacene.  Using these optimized structures, we evaluated
the adiabatic singlet-triplet energy gaps for the series and found that
the v2RDM-CASSCF gap converges to 7.8 \kcal in the limit of an
infinitely-long acene molecule; this estimate is larger than estimates
from other methods that include dynamical correlation effects, such as the
particle-particle random phase approximation.  We also demonstrated that
v2RDM-CASSCF predicts increasingly similar structures for the
lowest-energy singlet and triplet states as the length of the acene
molecules increases.  This similarity is a signature of the open-shell
nature of the singlet state.  We caution, however, that several recent
analyses\cite{Lee:2017:602,Dupuy:2018:134112} suggest that the limited
consideration of nondynamical correlation effects can lead to
qualitatively different results than more rigorous considerations.  In the
present context, a more rigorous description of the system might include
an expanded active space that incorporates some portion of the
$\sigma$-network.  Even then, a complete description of the system should
include dynamical correlation effects beyond those inadvertently captured
by large-active-space v2RDM-CASSCF.

Lastly, we note that the energies and analytic energy gradients obtained
from v2RDM-CASSCF are state specific, and the v2RDM-CASSCF procedure
itself is applicable only to the lowest-energy state of a given spin
symmetry.  In principle, a state-averaged-CASSCF-like method could be
achieved by evaluating excited states within the extended random phase
approximation
(ERPA),\cite{Rowe:1968:153,Chatterjee:2012:204109,vanAggelen:2013:50,DePrince:2016:164109}
and optimizing the orbitals for both the ground and excited states.  The
excited-state RDMs required for this procedure can be extracted from
ERPA-derived excited-state wave functions.

%Additions to our library of v2RDM-CASSCF methods in the Q-Chem software
%package include a GPU-accelerated implementation and inclusion of methods
%for including dynamic electron correlation.  [\textit{Not sure what to
%include here. It might be a good plans for featuring our plans for
%librdm.}]

%%%%%%%%%%%%%%%%%%%%%%%%%%%%%%%%%%%%%%%%%%%%%%%%%%%%%%%%%%%%%%%%%%%%%%
% Supporting information
%%%%%%%%%%%%%%%%%%%%%%%%%%%%%%%%%%%%%%%%%%%%%%%%%%%%%%%%%%%%%%%%%%%%%%
\vspace{0.3cm}
{\bf Supporting information.}  Harmonic vibrational frequencies computed
at the PQG/cc-pVDZ level of theory using ``loose'' and ``tight''
convergence criteria, equilibrium structures for the linear polyacene
series computed at the PQG/cc-pVDZ level of theory, and C--C bond lengths
for the long edge of the linear polyacene molecules up to dodecacene.

%%%%%%%%%%%%%%%%%%%%%%%%%%%%%%%%%%%%%%%%%%%%%%%%%%%%%%%%%%%%%%%%%%%%%%
% Acknowledgments
%%%%%%%%%%%%%%%%%%%%%%%%%%%%%%%%%%%%%%%%%%%%%%%%%%%%%%%%%%%%%%%%%%%%%%
\vspace{0.3cm}
{\bf Acknowledgments.} This material is based upon work supported by the
Army Research Office Small Business Technology Transfer (STTR) program
under Grant No. W911NF-16-C-0124.

%%%%%%%%%%%%%%%%%%%%%%%%%%%%%%%%%%%%%%%%%%%%%%%%%%%%%%%%%%%%%%%%%%%%%%
% Figures
%%%%%%%%%%%%%%%%%%%%%%%%%%%%%%%%%%%%%%%%%%%%%%%%%%%%%%%%%%%%%%%%%%%%%%
%\clearpage
%\input{figures}

%%%%%%%%%%%%%%%%%%%%%%%%%%%%%%%%%%%%%%%%%%%%%%%%%%%%%%%%%%%%%%%%%%%%%%
% Tables
%%%%%%%%%%%%%%%%%%%%%%%%%%%%%%%%%%%%%%%%%%%%%%%%%%%%%%%%%%%%%%%%%%%%%%
%\clearpage
%\input{tables}

%%%%%%%%%%%%%%%%%%%%%%%%%%%%%%%%%%%%%%%%%%%%%%%%%%%%%%%%%%%%%%%%%%%%%%
% Bibliography
%%%%%%%%%%%%%%%%%%%%%%%%%%%%%%%%%%%%%%%%%%%%%%%%%%%%%%%%%%%%%%%%%%%%%%
\clearpage
\bibliography{v2rdm_df_gradients.bib}

\end{document}